\documentclass[a4paper,11pt]{article}
\usepackage{jheppub} 

\usepackage[T1]{fontenc} 
\usepackage{xcolor}
\usepackage{caption}
\usepackage{subcaption}
\usepackage[normalem]{ulem}

\title{\boldmath Spread Complexity and Topological Transitions in the Kitaev Chain}

\author[a]{Pawel Caputa,}
\author[b]{Nitin Gupta,}
\author[b,c]{S. Shajidul Haque,}
\author[a]{Sinong Liu,}
\author[b,c]{Jeff Murugan,}
\author[b,c]{and Hendrik J.R. Van Zyl}

\affiliation[a]{Faculty of Physics, University of Warsaw, ul. Pasteura 5, 02-093 Warsaw, Poland}
\affiliation[b]{The Laboratory for Quantum Gravity \& Strings, Department of Mathematics \& Applied Mathematics, University of Cape Town, Cape Town, South Africa}
\affiliation[c]{The National Institute for Theoretical and Computational Sciences, Private Bag X1, Matieland, South Africa}

\abstract{A number of recent works have argued that quantum complexity, a well-known concept in computer science that has re-emerged recently in the context of the physics of black holes, may be used as an efficient probe of novel phenomena such as quantum chaos and even quantum phase transitions. In this article, we provide further support for the latter, using a 1-dimensional p-wave superconductor - the Kitaev chain - as a prototype of a system displaying a topological phase transition. The Hamiltonian of the Kitaev chain manifests two gapped phases of matter with fermion parity symmetry; a trivial strongly-coupled phase and a topologically non-trivial, weakly-coupled phase with Majorana zero-modes. We show that Krylov-complexity (or, more precisely, the associated spread-complexity) is able to distinguish between the two and provides a diagnostic of the quantum critical point that separates them. We also comment on some possible ambiguity in the existing literature on the sensitivity of different measures of complexity to topological phase transitions.}

\begin{document} 
\maketitle
\flushbottom

\bibliographystyle{JHEP}
\section{Introduction and motivation}
\label{sec:intro}
Nature is complex. Not {\it complicated}, like an aeroplane; {\it complex}, like a single human cell. This is why much of our time as theoretical physicists is spent searching for toy models; simple enough to solve but rich enough to contain some universal lessons about Nature. Not all toy models are created equal. Poor ones are simply trivial. Better ones answer the questions we set out to answer, but the best ones are those, like the harmonic oscillator or the Ising model, whose lessons go far beyond the problems they were conceived for. The Kitaev chain, a 1-dimensional tight-binding chain of spinless electrons supplemented with a superconducting term, falls into this last class. Originally introduced by Kitaev \cite{Kitaev_2001} in order to furnish free Majorana fermions that would facilitate topological quantum computing logic gates, this model has since matured into a versatile laboratory in which to test all manner of questions involving topological phases, critical points and quantum matter more generally \cite{Leijnse:2012tn}.\\

\noindent
The Kitaev Hamiltonian, $H_{\mathrm{Kitaev}}(J,\mu,\Delta)$, is parameterized by three real parameters; a tight-binding coupling $J$ that controls the nearest-neighbour hopping between sites on the 1-dimensional lattice, a chemical potential, $\mu$, and a p-wave pairing gap, $\Delta$ controlling the superconducting term. The Hamiltonian exhibits two distinct gapped phases; when $|\mu|>|J|$ the system finds itself in a {\it trivial phase} with a unique ground state independent of the boundary conditions, while in its {\it topological phase},  characterised by $|\mu|<|J|$, it manifests Majorana zero-modes in the open Kitaev chain, and odd fermion parity of the groundstate in the closed chain with periodic boundaries. For different values of its parameters, the Kitaev chain shares many properties with several other well-known quantum systems: 
\begin{itemize}
    \item When $J,\Delta > 0$, the Kitaev chain belongs to the same universality class as the Ising model. More precisely, the Kitaev and transverse Ising chains can be mapped into each other through a Jordan-Wigner transform \cite{Greiter_2014}.
    \item The Su-Schrieffer-Heeger (SSH) model of polyacetylene \cite{Su:1979ua}, is a 1-dimensional dimerized lattice of electrons in which each unit cell consists of two sites $A$ and $B$ and a hopping term connecting the two sub-lattices. Performing a particle-hole transformation, $c_{B,n}\to c_{B,n}^{\dagger}$ on all electrons living on the $B$ sublattice, converts it, for specific parameter values, into the p-wave superconductor, and  
    \item Turning off the superconducting term ($\Delta=0$)  $H_{\mathrm{Kitaev}}\to H_{\mathrm{XX}}$, the isotropic limit of the XY model \cite{Franchini:2016cxs}.
\end{itemize}
\noindent
In most cases these, and other, properties derive from a study of the spectrum of eigenvalues, either analytically or numerically, of the Hamiltonian matrix. However, the explicit diagonalization of an $N\times N$ matrix is a computationally expensive exercise for large $N$. It behoves us then to search for more bespoke tools to study such many-body systems, especially in the presence of further complications such as interactions, disorder and large system sizes. One particularly exciting set of such tools comes from the growing influence of quantum information theory in both condensed matter and high energy physics. Included in this set are concepts such as Von Neumann entropy, entanglement spectra, Holevo entropy, quantum error correction and, of particular interest to us, quantum complexity \cite{nielsen00,Watrous_2008}.
\\

\noindent
In a nutshell, quantum complexity - the quantum analog of the well-known computer science concept of computational complexity - attempts to quantify, as efficiently as possible, how difficult it is to synthesize a particular quantum state using a set of allowed unitary operations. A substantial amount of the recent interest in quantum complexity in high energy physics is motivated by developments in the understanding of the physics of black hole information loss. In particular, the AdS/CFT correspondence \cite{Maldacena:1997re} that connects quantum gravity in $D$-dimensional anti-de Sitter space to a conformal field theory living on its $(D-1)$-dimensional boundary, also connects quantum complexity in the field theory to various quantities in the gravity side of the correspondence via the so-called complexity/volume and complexity/action conjectures \cite{Brown:2015bva}. Indeed, complexity as a resource in quantum information \cite{Halpern:2021ufe}, has featured prominently in discussions on black hole firewall \cite{Harlow:2013tf}, fast scrambling near black hole horizons \cite{Aaronson:2016vto, Susskind:2014rva, Stanford:2014jda, Brown:2015bva}, quantum chaos \cite{Maldacena:2015waa} and the properties of the holographic dictionary itself \cite{Bouland:2019pvu}.  \\

\noindent
Forgetting about the connection to black holes and gravitational physics for the moment, one, not insignificant problem is that, while the central question of finding the smallest number of unitary operations  that connect a fixed reference and target state remains, the actual computation of quantum complexity is not unique. There are by now several contenders (see e.g. reviews \cite{Chapman:2021jbh,Chen:2021lnq} and references therein). Among these, the Nielsen approach \cite{Nielsen1,Nielsen_2006, Nielsen3} phrases the problem as one of finding the minimal geodesics connecting two points on the manifold of quantum states.  The inputs required for its computation are the desired reference and target state along with the choices of unitary gates, their penalties and a cost function. These ambiguities make it difficult to determine definitively which features of a quantum system the complexity is sensitive to. That said, Nielsen complexity has been shown already to be a useful probe of topological phase transitions in \cite{Ali:2018aon, Liu:2019aji, Xiong:2019xoh} with some variation between the different studies, possibly due to the ambiguities outlined above\footnote{It must be pointed out though that the authors of \cite{Liu:2019aji, Xiong:2019xoh} have argued that appropriate choices for the ambiguities may detect topological phase transitions as discontinuities in Nielsen complexity}.\\

\noindent
  More recently, a proper subset of us (SL \& PC) have argued in \cite{Caputa:2022eye} that a recently introduced measure of state complexity called \textit{spread complexity} \cite{Balasubramanian:2022tpr} is able to disambiguate between the topologically distinct phases of the SSH model \cite{Su:1979ua}. Unlike Nielsen complexity however, spread complexity requires as input only the desired reference and target state along with the Hamiltonian of the system in question making it less susceptible to the ambiguities of Nielsen complexity. As a variant of Krylov complexity, it also comes equipped with a clear algorithmic construction, making it computationally efficient. Taken together, these properties make it a useful computational tool to study various aspects of many-body quantum systems. For example,  in addition to successfully probing the topological phase structure of the SSH model, spread complexity has also been been shown to be a useful diagnostic of quantum chaos \cite{Balasubramanian:2022tpr} to supplement other diagnostics \cite{Ali:2019zcj,Bhattacharyya:2019txx,Bhattacharyya:2020art,Bhattacharyya:2020iic} such as the out-of-time-order correlators (OTOCs) \cite{Larkin1969QuasiclassicalMI} or the spectral form factor (SFF) \cite{Dyer:2016pou,Liu:2018hlr}. This article is an extension of these ideas and provides further substantiation of spread complexity as a good probe of quantum phase transitions. \\

\noindent
 The rest of this article is organised as follows: Since the spread complexity that is the focus of this paper is closely related to K(rylov)-complexity, we begin the next section with a brief self-contained review of spread complexity based on the Krylov basis \cite{Balasubramanian:2022tpr}. Having reviewed the technology, in section 3 we diagonalize the Kitaev Hamiltonian in terms of $SU(2)$ coherent states and subsequently compute the spread complexity of the 1-dimensional Kitaev chain. This is followed by a detailed study of the spread complexity of the p-wave superconductor for three different circuits, involving different choices of target state.    Finally we conclude and discuss future directions.
Appendix A gives a self-contained treatment of the $SU(2)$ coherent states \cite{Perelomov}, \cite{Gazeau:2009zz} utilized in the computation of the spread complexity. Finally, in Appendix B we apply the methods developed in \cite{Caputa:2022eye} to a generalization of the Kitaev chain discussed above, with the open/closed boundary conditions replaced by {\it twisted boundary conditions} \cite{Kawabata:2017zsb}. The phase that enters the problem with the twisting, together with the p-wave pairing gap parameter $\Delta$ endow the Kitaev chain further structure beyond that of the SSH model. We provide some preliminary evidence that some of this structure manifests in the spread complexity. \\
\section{Krylov and Spread Complexity}
\label{sec:krylov}

In the interests of self-containment, we provide here a brief summary of the notion of spread complexity introduced in \cite{Balasubramanian:2022tpr} as well as the Krylov basis, both of which are central to the computations performed in this  paper. This section borrows heavily from a similar discussion in \cite{Caputa:2022eye}. \\ \\
\noindent
Our starting point is a general quantum state $|\Psi(s)\rangle$, related to some initial state $|\Psi_{0}\rangle$ by a unitary transformation
\begin{equation}
|\Psi(s)\rangle = e^{- i s H} |\Psi_0\rangle\,.\label{CIRQ1}
\end{equation}
We will refer to the state $|\Psi_0\rangle$ as the {\it reference state} and $|\Psi(s)\rangle$ as the {\it target state} of a unitary circuit with circuit Hamiltonian, $H$. The parameter $s$ is a circuit parameter that we will usually take to run from $0$ to $1$ as we traverse the circuit.  A useful measure of quantum state complexity can be defined as the spread of $|\Psi(s)\rangle$ in the Hilbert space \cite{Balasubramanian:2022tpr}, i.e., the spread of $|\Psi_0\rangle$ under the {\it unitary evolution} \eqref{CIRQ1}.  To quantify this spread, the target state may be expanded in some basis $\mathcal{B} = \left\{ |B_n\rangle, n=0,1,2, \cdots \right\}$ where we take $|B_0\rangle = |\Psi_0\rangle$, the reference state.  Then, we can define a cost function,
\begin{equation}
C_{\mathcal{B}}(s) = \sum_{n}  n |\langle \Psi(s) | B_n\rangle|^2\,,\label{CostKr}
\end{equation}
which is to be minimized over all choices of basis $\mathcal{B}$. We then define the spread complexity as
\begin{equation}
\mathcal{C}(s) = \textnormal{min}_{\mathcal{B}} \ C_{\mathcal{B}}(s),
\end{equation}
and the actual value of complexity corresponding to our target state is $\mathcal{C}(s=1)$.\\ \\
\noindent
The basis that satisfies the minimisation condition of the cost function \eqref{CostKr} is the Krylov basis \cite{Balasubramanian:2022tpr}.  The Krylov basis $\{|K_n\rangle\}$ can in turn be obtained by performing a Gram-Schmidt orthogonalization on the states $|O_n\rangle = H^n |\Psi_0\rangle$.  This is known as the Lanczos algorithm \cite{Lanczos1950AnIM} (see review \cite{viswanath2008recursion}).  Note that the Krylov basis depends on the choice of circuit Hamiltonian as well as the choice of reference state.  The circuit Hamiltonian is generally tri-diagonal in this basis in the sense that
\begin{equation}
H|K_n\rangle = a_n |K_n\rangle + b_n |K_{n-1}\rangle + b_{n+1} |K_{n+1}\rangle\,,
\end{equation}
where the so-called Lanczos coefficients $a_n$ and $b_n$ may be recursively constructed using the Lanczos algorithm.  The information captured by these coefficients is also contained in the moments of the return-amplitude
\begin{equation}
S(s) \equiv \langle \Psi(s) | \Psi(0)\rangle=\langle \Psi_0|e^{isH} | \Psi_0\rangle,
\end{equation}
that is the probability amplitude for the state to remain in the initial state. In many-body literature it is also often called the survival or Loschmidt amplitude \cite{Heyl:2017blm}.

Then, the target state may be expanded in the Krylov basis
\begin{equation}
|\Psi(s) \rangle = \sum_{n} \psi_n(s) |K_n\rangle\,,
\end{equation}
where, by construction, the coefficients satisfy the discrete Schrodinger equation
\begin{equation}
i \partial_s\psi_n(s) = a_n \psi_n(s) + b_n \psi_{n-1}(s) + b_{n+1}\psi_{n+1}(s)\,.
\end{equation}
With the knowledge of the Lanczos coefficients this equation may be solved with initial condition $\psi_n(0) = \delta_{n,0}$. Finally, in the Krylov basis the spread complexity now becomes
\begin{eqnarray}
\mathcal{C}(s) & = & \sum_{n} n |\psi_n(s)|^2 \,.\label{SCKr}
\end{eqnarray}
This definition generalises Krylov complexity (K-complexity for short) of the operator growth defined in \cite{Parker:2018yvk} (see also \cite{Barbon:2019wsy,Magan:2020iac,Rabinovici:2020ryf,Jian:2020qpp,Yin:2020oze,Dymarsky:2019elm,Dymarsky:2021bjq,Caputa:2021ori,Carrega:2020jrk,Kim:2021okd,Kar:2021nbm,Rabinovici:2022beu,Trigueros:2021rwj,Rabinovici:2021qqt,Hornedal:2022pkc,Bhattacharjee:2022vlt,Fan:2022xaa,Adhikari:2022whf,Muck:2022xfc,Patramanis:2021lkx,Caputa:2021sib, Bhattacharjee:2022qjw} and references therein) \footnote{There, by appropriate choice of the inner-product in the space of operators one can set $a_n=0$.}.
It is often useful to introduce a "complexity operator" by relations
\begin{equation}
\hat{K}  =  \sum_n n |K_n\rangle \langle K_n |,\qquad \mathcal{C}(s)\equiv \langle \Psi(s) | \hat{K}   | \Psi(s)\rangle \,.\label{SCKOp}
\end{equation}
These two relations (\eqref{SCKr} and \eqref{SCKOp}) will be our working definitions of the spread complexity for the remaining parts of this work.

In principle, every quantum dynamics can be formulated in terms of this one-dimensional chain-like picture. In practice, for generic setups, extracting Lanczos coefficients and solving the Schrodinger equation may be only accessible numerically. Nevertheless, as it will also be the case in this work, when the circuit Hamiltonian is an element of a low-rank semi-simple Lie algebra, the Krylov basis and complexity may be straightforwardly extracted following the construction reviewed in \cite{Caputa:2021sib}.  In Appendix (\ref{su2Appendix}) we also show how the spread complexity may be computed directly from the relevant coherent state overlap.  

Last but not the least, as in \cite{Caputa:2022eye}, we will employ a natural extension of the above definitions to free theories. Namely, the states of our interest will have the product form for each momentum mode. Similarly, to geometric definitions (Nielsen type) we will then compute the spread complexity for each momentum separately and then just sum the answers (or integrate in the continuum limit)\footnote{We should admit that, for general states, demonstrating that this prescription is equivalent to the canonical definition of K-complexity in position space or the definition starting from the return amplitude is not immediately obvious. In this work we will then follow the steps outlined above and leave studies of relations between different prescriptions as an important and interesting future direction. }.

\section{The Kitaev chain and topological phase transitions}
\label{sec:top}
In this section we review our main physical setup of the Kitaev model \cite{Kitaev_2001}. The Kitaev chain Hamiltonian is given by 

\begin{equation}
H = \sum_{j = 1}^L \left[ -\frac{J}{2}  \left( c_j^\dag c_{j+1} + c_{j+1}^\dag c_j  \right)  - \mu \left( c_j^\dag c_j - \frac{1}{2}   \right)   +   \frac{1}{2}\left( \Delta c_j^\dag c_{j+1}^\dag +\Delta^* c_{j+1} c_j  \right)   \right]\,,   \label{KiteavChainH}
\end{equation}

\noindent
where the $c_i$ ($c_{i}^{\dagger}$) are spinless fermionic annihilation (creation) operators acting at site $i$ on an $L$-site lattice,  defined by the algebra,
\begin{equation}
\left\{ c_j^\dag, c_k\right\} = \delta_{j k}. 
\end{equation}
The hopping amplitude, $J$ and chemical potential $\mu$ are real parameters while the superconducting pairing strength, $\Delta$, is generally complex, although it can always be chosen to be a positive real number by rescaling the annihilation operator $c_i \to e^{i \frac{1}{2}\operatorname{arg} \Delta} c_i $. In what follows, we consider $\Delta$ real which is equivalent to a rescaling factor of either $e^{i \frac{1}{2}\operatorname{arg} \Delta} = \pm 1$ or $e^{i \frac{1}{2}\operatorname{arg} \Delta} = \pm i$, for positive or negative $\Delta$, respectively. The model exhibits a topological phase transition when the quasi-particle spectrum is gapless \cite{Kitaev_2001}.   In the continuum limit, $L \rightarrow \infty$, this corresponds to a topological phase transition at $|\frac{\mu}{J}| = 1$ with a topological phase in the regime $|\frac{\mu}{J}| < 1$ and a non-topological phase in the regime $|\frac{\mu}{J}| > 1$. Furthermore, the winding number of the topological phase is determined by the sign of $\Delta$, which implies that there are two topological phases in the $\frac{\Delta}{J}-\frac{\mu}{J}$ plane, separated by $\Delta =0$.\\ 

To facilitate a discussion of the spectral properties of the model, it is convenient to perform a Fourier transform
\begin{equation}
c_j  = \frac{1}{\sqrt{L}}  \sum_{n} e^{i k_n j} a_{k_n}\,, 
\end{equation}
which puts the Hamiltonian into the form

\begin{equation}
\begin{split}
H =& \sum_{n=0}^{L-1}\left[-(\mu + J \cos(k_n) ) \left(a^\dag_{k_n} a_{k_n} - \frac{1}{2} \right) +   \frac{1}{2} \Delta \left( e^{-i k_n} a^\dag_{-k_n} a^\dag_{k_n} + e^{i k_n}a_{k_n} a_{-k_n}   \right)   \right]  \\
=& \sum_{n=0}^{\lceil \frac{L-1}{2} \rceil }\left[-(\mu + J \cos(k_n) ) \left(a^\dag_{k_n} a_{k_n}  - a_{-k_n} a^\dag_{-k_n} \right)+   \Delta \left( e^{-i k_n} a^\dag_{-k_n} a^\dag_{k_n} + e^{i k_n}a_{k_n} a_{-k_n}   \right)   \right].
\label{FourierH}
\end{split}
\end{equation}
The values of $k_n$ are fixed by requiring that the momentum-space creation and annihilation operators satisfy the correct anti-commutation relations as well as imposing the desired boundary condition on $c_j$.  The generalized boundary condition, 
\begin{equation}
c_{L+1} = e^{i \phi} c_1,
\end{equation}
with $e^{i \phi} = 1 \textnormal{ and } e^{i \phi} = -1$ denoting periodic and anti-periodic boundary conditions respectively, corresponds to the set of momenta
\begin{equation}
k_n = \frac{2 \pi}{L}\left(  -\left\lceil \frac{L-1}{2} \right\rceil  +  n + \frac{\phi}{2 \pi}  \right) \ \ \ ; \ \ \ n = 0, 1, 2, \cdots L-1\,.    \label{kVals}
\end{equation}
For even $L$, the Kitaev chain with the anti-periodic boundary condition (APBC)  may be written as

\begin{equation}
H = -\sum_{k_n > 0} \left[ 2 (\mu + J \cos(k_n) ) J_{0}^{(k_n)} - i \Delta \sin(k_n) \left(J_{+}^{(k_n)} - J_{-}^{(k_n)}\right)  \right].   \label{Hsu2Basis}
\end{equation}
Closely related expressions for periodic boundary conditions and for odd $L$ can be found in Appendix A of \cite{Kawabata:2017zsb}. In this form of the Hamiltonian we defined the ladder operators
\begin{eqnarray}
J_{0}^{(k_n)} & \equiv &  \frac{1}{2}\left( a^\dag_{k_n} a_{k_n} - a_{-k_n} a^\dag_{-k_n}  \right)\,,    \nonumber \\
J_{+}^{(k_n)} & \equiv & a^\dag_{k_n} a^\dag_{-k_n}\,, \label{su2Generators} \\
J_{-}^{(k_n)} & \equiv & a_{-k_n} a_{k_n}\,,  \nonumber
\end{eqnarray}
that satisfy the $su(2)$ algebra
\begin{eqnarray}
\left[ J_{0}^{(k_n)}, J_{\pm}^{(k_{n'})}  \right] & = & \pm \delta_{n, n'}  J_{\pm}^{(k_{n})}\,,   \nonumber \\
\left[ J_{+}^{(k_n)}, J_{-}^{(k_{n'})}  \right] & = & 2 \delta_{n, n'}  J_{0}^{(k_n)}\,. \nonumber
\end{eqnarray}
A key point to note is that, after defining, 
\begin{equation}
\vec{R}(k_n) \equiv \begin{pmatrix}
R_1 \\
R_2 \\
R_3 \\
\end{pmatrix}=
\begin{pmatrix}
0 \\
\Delta \sin k_n \\
\mu+ J \cos k_n \\
\end{pmatrix},
\end{equation}
the Hamiltonian (\ref{Hsu2Basis}) is essentially of the same form 
\begin{equation}
H = -\sum_{k_n>0} \left[ 2 R_3 J_0^{(k_n)} - i R_2 (J_{+}^{(k_n)} - J_{-}^{(k_n)})   \right],  \label{su2KitaevH}
\end{equation}

\noindent
as that studied in \cite{Caputa:2022eye}, with the only differences being the values of the coefficients $R_2, R_3$ as well as only summing over positive values of $k$.  The latter feature may be anticipated from the map between the SSH Hamiltonian and the periodic Kitaev chain \cite{Bandyopadhyay:2021dei}. \\

\noindent
The full excitation spectrum of the Hamiltonian may be represented in terms of $SU(2)$ coherent states\footnote{For each $k$ the states  $|\psi_{+}\rangle = \cos(\frac{\varphi_{k}}{2}) e^{- i \tan(\frac{\varphi_{k}}{2}) J_{+}^{(k)} } | \frac{1}{2}, -\frac{1}{2}\rangle$ and $|\psi_{-}\rangle = \sin(\frac{\varphi_{k}}{2}) e^{i \cot(\frac{\varphi_{k}}{2}) J_{+}^{(k)} } | \frac{1}{2}, -\frac{1}{2}\rangle$ are eigenstates with eigenvalues $\epsilon_{\pm} = \pm  \sqrt{(\mu + J \cos \frac{\varphi_k}{2} )^2 + \left( \Delta \sin \frac{\varphi_k}{2} \right)^2   }$.  }  and, in particular, the ground state is given by 
\begin{equation}
|\Omega\rangle = \prod_{k_n > 0} \sin \left|\frac{\varphi_{k_n}}{2}\right| \exp \left[  e^{-i \psi_{k_n}} \cot \frac{\varphi_{k_n}}{2} J_{+}^{(k_n)} \right] \left| \frac{1}{2}, -\frac{1}{2}\right\rangle_{k_n}\,,    
\label{grdstt}
\end{equation}
where we have introduced the spherical coordinates $(R(k_n),\varphi_{k_n},\psi_{k_n})$ to rewrite the parameter vector $\vec{R}(k_n)$. In particular,
\begin{equation}
\left\{
\begin{matrix}
R(k_n) = \left|\vec{R}(k_n) \right| \\
\varphi_{k_n} = \arctan \frac{|\Delta | \sin k_n}{\mu + J \cos k_n}
\\
\psi_{k_n} = \frac{\pi}{2}-\operatorname{arg} \Delta  \\
\end{matrix}
\right., \qquad k_n >0\,.
\end{equation}
Note that these relations remain true for complex $\Delta$ as well. To compare our conventions with \cite{Liu:2019aji}, set
\begin{equation}
\nonumber
\theta_{k_n} = \operatorname{sgn} \Delta \cdot \frac{\pi-\varphi_{k_n}}{2}\,.
\end{equation}

\noindent
The reference state above is a tensor product of the lowest weight $su(2)$ states in the spin-$\frac{1}{2}$ representation, for each positive value of $k_n$. \\

\noindent
Before proceeding, a few points are in order.
\begin{itemize}                      
\item The first is a comment on some of the existing literature on this topic. The periodic Kitaev chain with even $L$ and both short- and long-range pairing has been studied in  \cite{Liu:2019aji} where it was determined that Nielsen's approach to circuit complexity may be used to detect the topological phase transition.  In this paper we will instead study the spread complexity to determine whether it is also sensitive to this transition.  Another model exhibiting a topological phase transition, the SSH model \cite{Su:1979ua}, has been studied using both circuit complexity \cite{Ali:2018aon} and spread complexity \cite{Caputa:2022eye}. Taken together, these results seem to point to spread complexity being sensitive to the topological phase transition, while Nielsen complexity is not. This would appear to be in contrast to our observations of the topological transition in the Kitaev chain. There are a few possible explanations of this, the simplest of which is that in \cite{Ali:2018aon} the amplitude of the fluctuations in the early time regime, which is essential to detect the transition as argued in \cite{Caputa:2022eye}, was not probed carefully enough.
\item
The second concerns a generalization of the Kitaev chain to account for twisted boundary conditions. The associated Hamiltonian and site-dependent chemical potential are
\cite{Kawabata:2017zsb}
\begin{eqnarray}
H & = & \sum_{j = 1}^{L-1} \left[ -\frac{J}{2}  \left( c_j^\dag c_{j+1} + c_{j+1}^\dag c_j  \right)     +   \frac{\Delta}{2}\left( c_j c_{j+1} + c_{j+1}^\dag c_j^\dag  \right)   \right]   - \sum_{j=1}^L \mu_j \left( c_j^\dag c_j - \frac{1}{2}   \right) \nonumber \\
  & & +  b \left[ -\frac{J}{2}(e^{i \phi_1} c_L^\dag c_1 + e^{-i \phi_1} c_1^\dag c_L) +   \frac{\Delta}{2}(e^{i \phi_2} c_L c_1 + e^{-i \phi_2} c_1^\dag c_L^\dag )   \right],\nonumber \\
	\mu_j & =& \mu + \mu (a - 1)(\delta_{j,1} + \delta_{j,L}).   \label{TwistedKitaevH}
\end{eqnarray}
The bulk chemical potential, hopping amplitude and superconducting pairing strength are as before, while the chemical potential at the boundary sites $i=1,L$ has a modified form $\mu_1 = \mu_L = a \mu$.  In addition, the terms connecting sites $1$ and $L$ pick up additional phases $\phi_1, \phi_2$.  These may run over an interval of $2\pi$ and we will find it convenient to parameterise the interval as $\phi_i \in [-\pi, \pi)$. \\

\noindent
As usual, the model exhibits a topological phase transition when the ground state energy becomes degenerate.  As shown in \cite{Kawabata:2017zsb}, this coincides with the existence of Majorana zero modes.  Such zero modes are not present for arbitrary choices of the system parameters.  This model should thus allow one to assess spread complexity as a probe of the topological phase transition.  Essentially, one would want a probe to diagnose the phase transition if and only if it is present. To perform the computation one would have to obtain, numerically or otherwise, the Krylov basis.  We leave this detailed study to future work but provide some promising approximate results in this direction in Appendix \ref{twisted-appendix}. 
\end{itemize}

\section{Spread complexity as a probe of topological phase transitions}
\label{sec:k-comp}

As previously discussed, eigenstates of the Kitaev Hamiltonian (\ref{su2KitaevH}) may be fully described in terms of $SU(2)$ coherent states.  In addition, spread complexity is a function of the Hamiltonian parameters and the choice of reference and target state which determine the circuit.  In what follows, we consider circuits connecting the ground state of the Kitaev chain (taken as the target state) to a number of natural and universally definable choices of reference states. Following \cite{Caputa:2022eye} we compute the spread complexity for each of these circuits and show its  sensitivity to the topological phase transition in all cases. In this sense, this demonstrates the robustness of spread complexity as a probe of topological phase transitions, at least for the Kitaev chain and the models related to it. Unless otherwise stated we work with the Hamiltonian (\ref{Hsu2Basis}) valid for even $L$ and anti-periodic boundary conditions. 

\subsection{Circuit 1} \label{pkccHm}
All the circuits considered here and below will take the Kitaev chain ground state as the target state and we distinguish between the different circuits by their choice of reference state.
For convenience, let us rewrite the ground state \eqref{grdstt} utilising the Baker-Campbell-Hausdorff identity, as
\begin{eqnarray}
|\Omega(s=1) \rangle &=& \bigotimes_{k>0} |\Omega_k(s=1) \rangle, \\
|\Omega_k(s=1)\rangle &=&   \exp \left[ \frac{\pi-\varphi_k}{2} e^{-i \left(\frac{\pi}{2} -\operatorname{arg} \Delta \right)} J_+^{(k)} -\frac{\pi-\varphi_k}{2} e^{i \left(\frac{\pi}{2} -\operatorname{arg} \Delta \right)} J_-^{(k)} \right] \left|\frac{1}{2}, -\frac{1}{2} \right\rangle_k, \nonumber \\
&=& \exp \left[ -i \frac{\pi-\varphi_k}{2} \left( e^{i \operatorname{arg} \Delta } J_+^{(k)} + e^{-i  \operatorname{arg} \Delta } J_-^{(k)}\right) \right] \left|\frac{1}{2}, -\frac{1}{2} \right\rangle_k.
\end{eqnarray}
As a first example of reference state, we will take
\begin{equation}
|\Omega_k(s=0)\rangle = \left| \frac{1}{2}, \ \operatorname{sgn}\left(\mu+ J \cos k \right) \cdot \frac{1}{2} \right\rangle_k.
\label{gsM}
\end{equation}
In a sense, this reference state is a ``natural'' one to choose, since it is nothing but the ground state of the free fermionic Hamiltonian (or the mass part of the Hamiltonian) in \eqref{Hsu2Basis}, i.e., it is the ground state of
\begin{equation}
H_M =  -\sum_{k>0}  2 R_3 J_0^{(k)} = - \sum_{k>0} 2|R_3| \left( \operatorname{sgn} R_3 \cdot J_0^{(k)}\right),
\end{equation}
with $R_3 = \mu+J \cos k$ and where in the last expression above, we have expressed the Hamiltonian in a form that makes the ground state more obvious.
Note that \eqref{gsM} is equivalent to
\begin{equation}
|\Omega_k(s=0)\rangle = \exp \left[ -i \frac{\pi}{2} \Theta \left(\mu+ J \cos k \right) \cdot \left(  J_+^{(k)} +  J_-^{(k)}\right)  \right] \left|\frac{1}{2}, -\frac{1}{2} \right\rangle_k,
\end{equation}
where we have introduced the step function $\Theta$. 

The first circuit we consider, connecting the free fermion ground state to the Kitaev chain ground state, thus takes the form
\begin{equation}
|\Omega_k(s)\rangle = \exp \left[ -i s \left( \frac{\pi -\varphi_k }{2}\operatorname{sgn}\Delta -\frac{\pi}{2} \Theta \left(\mu+ J \cos k \right) \right) \left(  J_+^{(k)} +  J_-^{(k)}\right) \right] |\Omega_k(s=0)\rangle.
\end{equation}
The spread complexity is then given by 
\begin{eqnarray}
\mathcal{C}_k(s=1) &=& \sin^2 \left( \theta_k -\frac{\pi}{2} \Theta \left(\mu+ J \cos k \right) \right) = \frac{1}{2} \left(1 + \operatorname{sgn}\left(\mu+ J \cos k \right)  \cdot  \cos 2\theta_k \right) \nonumber \\
&=& \frac{1}{2} \left(1 - \operatorname{sgn}\left(\mu+ J \cos k \right)  \cdot  \cos \varphi_k \right) \nonumber \\
&=& \frac{1}{2}\left(1 - \frac{|\mu + J \cos(k)|}{\sqrt{(\mu + J \cos(k) )^2 + \Delta^2 \sin^2(k)  }}\right).   \label{CCircuit1}
\end{eqnarray}
To find the total spread complexity we have to sum over the discrete values of $k_n$ for finite $L$, which becomes an integral in the continuum limit
\begin{equation}
C(J, \mu, \Delta) = \frac{1}{L} \sum_{n > 0} C_{k_n}(s=1) \rightarrow \frac{1}{\pi} \int_0^\pi dk\, C_{k}(s=1)\,.
\label{intCk}
\end{equation} 
\begin{figure}
    \centering
    \hspace{-1.0cm}
    \includegraphics[width=0.70\textwidth]{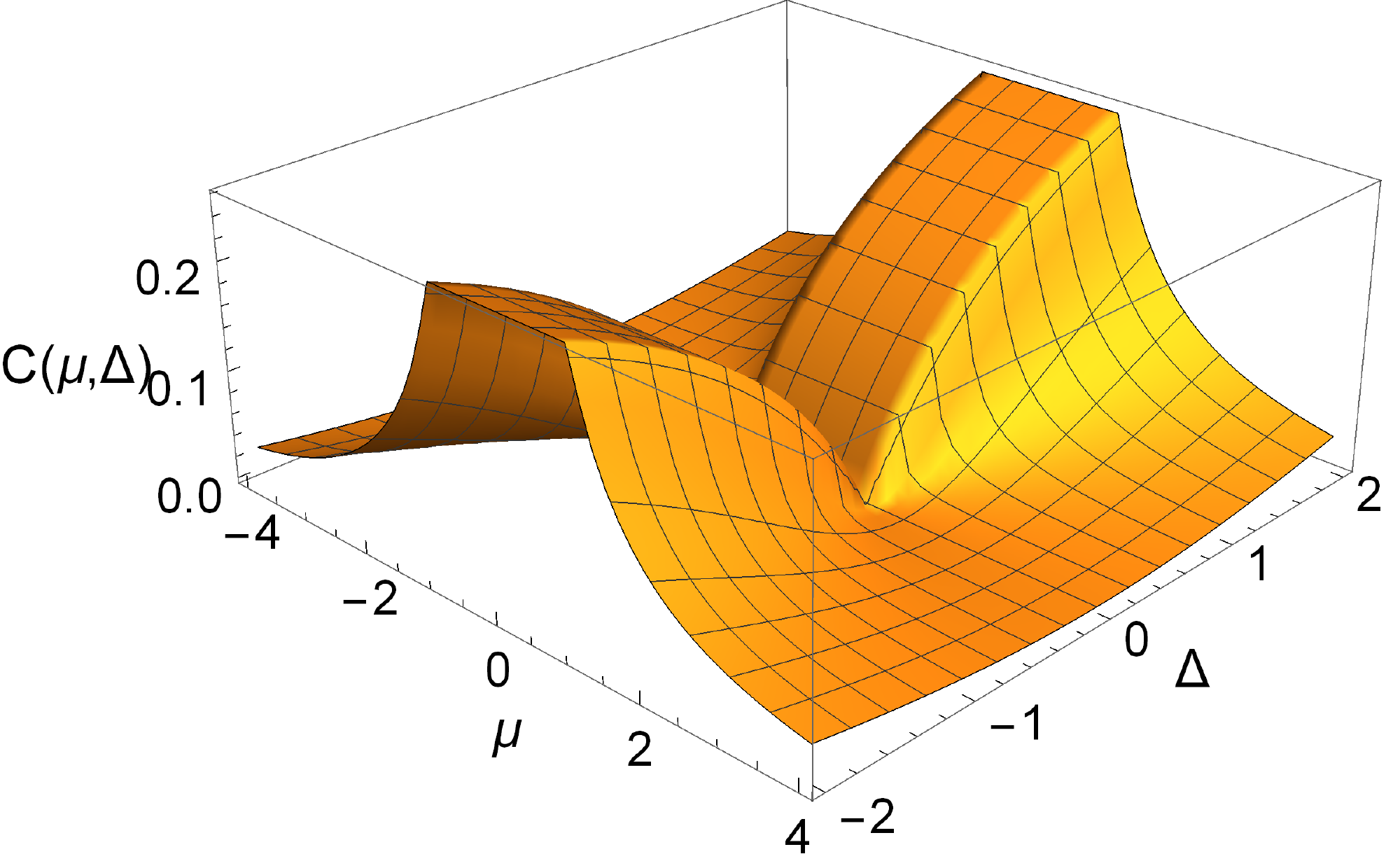}
    \caption{The spread complexity in the continuum limit for the circuit connecting the free fermion ground state to the Kitaev chain ground state. We have chosen $J=1$.  When $|\mu| < 1 $ the system is in the topological phase and the spread complexity is a $\Delta$-dependent constant.}
    \label{fig:Circuit13d}
\end{figure}

\begin{figure}[!h]
\begin{subfigure}{.5\textwidth}
	\centering
		\includegraphics[width=0.95\textwidth]{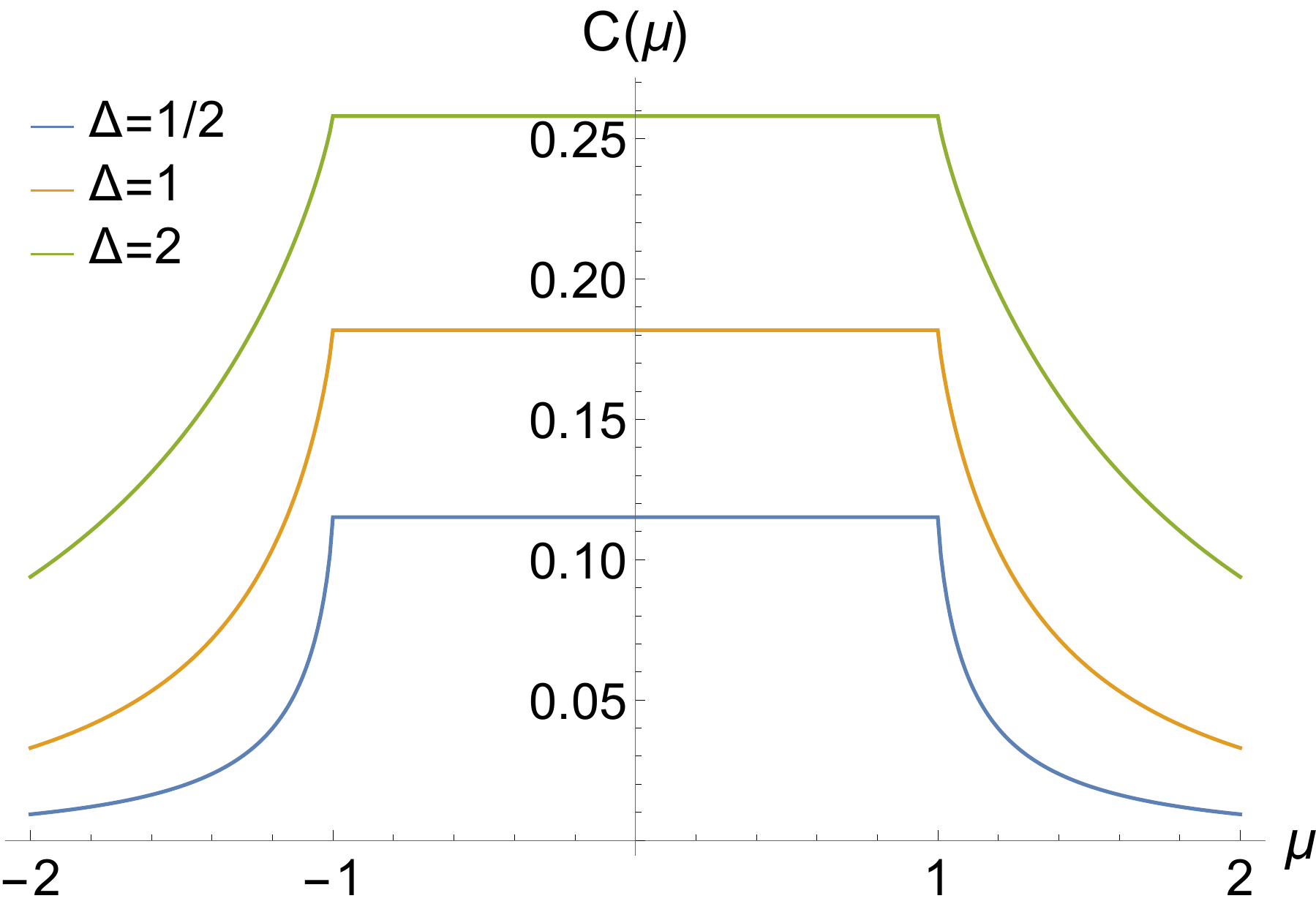}
		\caption{}
	\label{fig:InfiniteLCircuit1Deltas}
	\end{subfigure}
	\begin{subfigure}{.5\textwidth}
		\centering
		\includegraphics[width=0.95\textwidth]{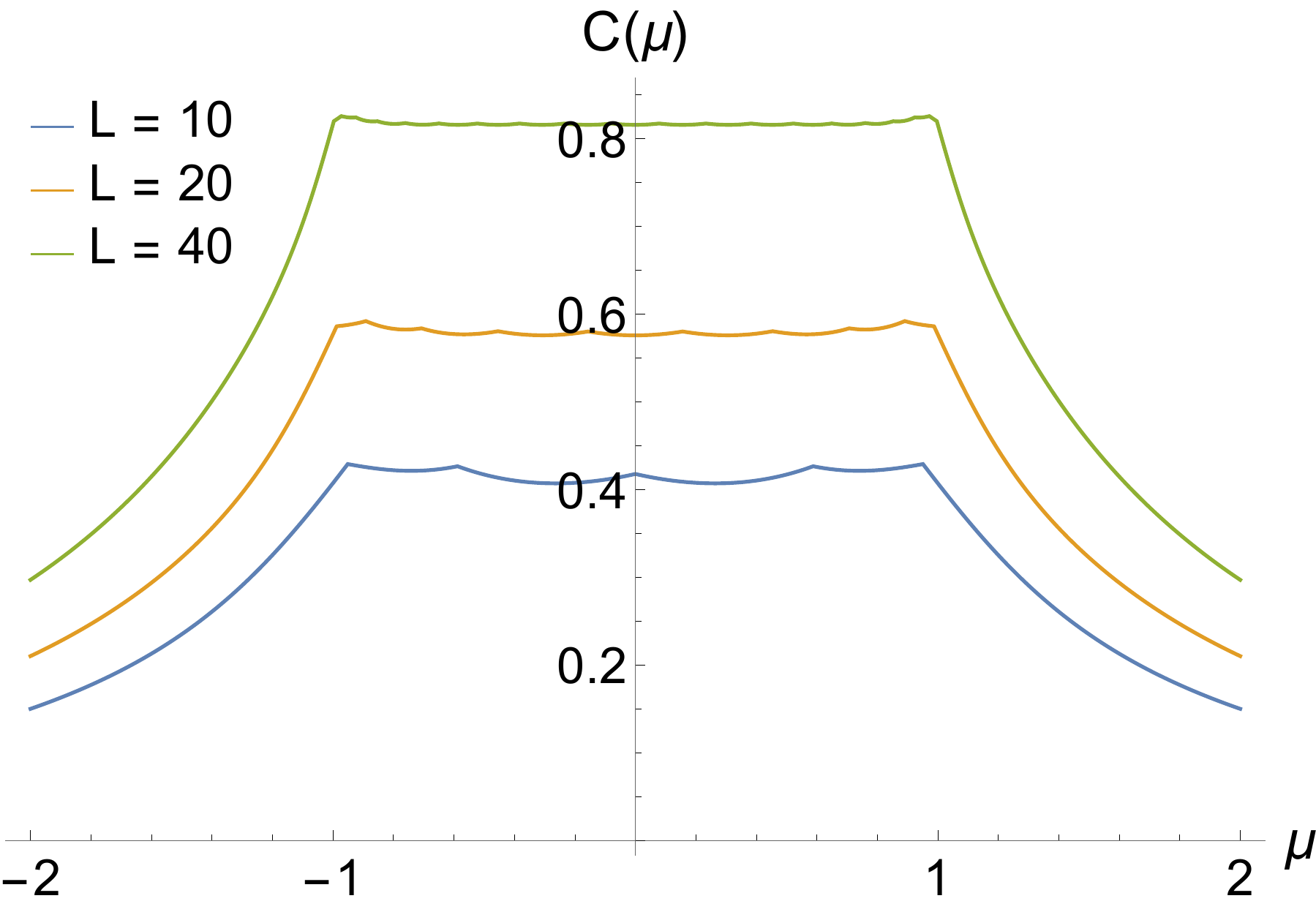}
		\caption{}
	\label{fig:FiniteLCircuit1}
	\end{subfigure}
		\caption{a) The spread complexity for the circuit connecting the free fermion ground state and Kitaev chain ground state as a function of the chemical potential, with $J =1$, and in the continuum limit.  The complexity is constant in the region $|\mu| <1$.  b) The spread complexity for the circuit connecting the free fermion ground state and Kitaev chain ground state as a function of the chemical potential, with $J =1$ and $\Delta=2$. The spread complexity has been divided by $\sqrt{L}$ for display purposes.  The top curve corresponds to $L=40$, the middle $L=20$ and the bottom $L=10$.}
\end{figure}

\begin{figure}[h]
	\centering
		\includegraphics[width=0.70\textwidth]{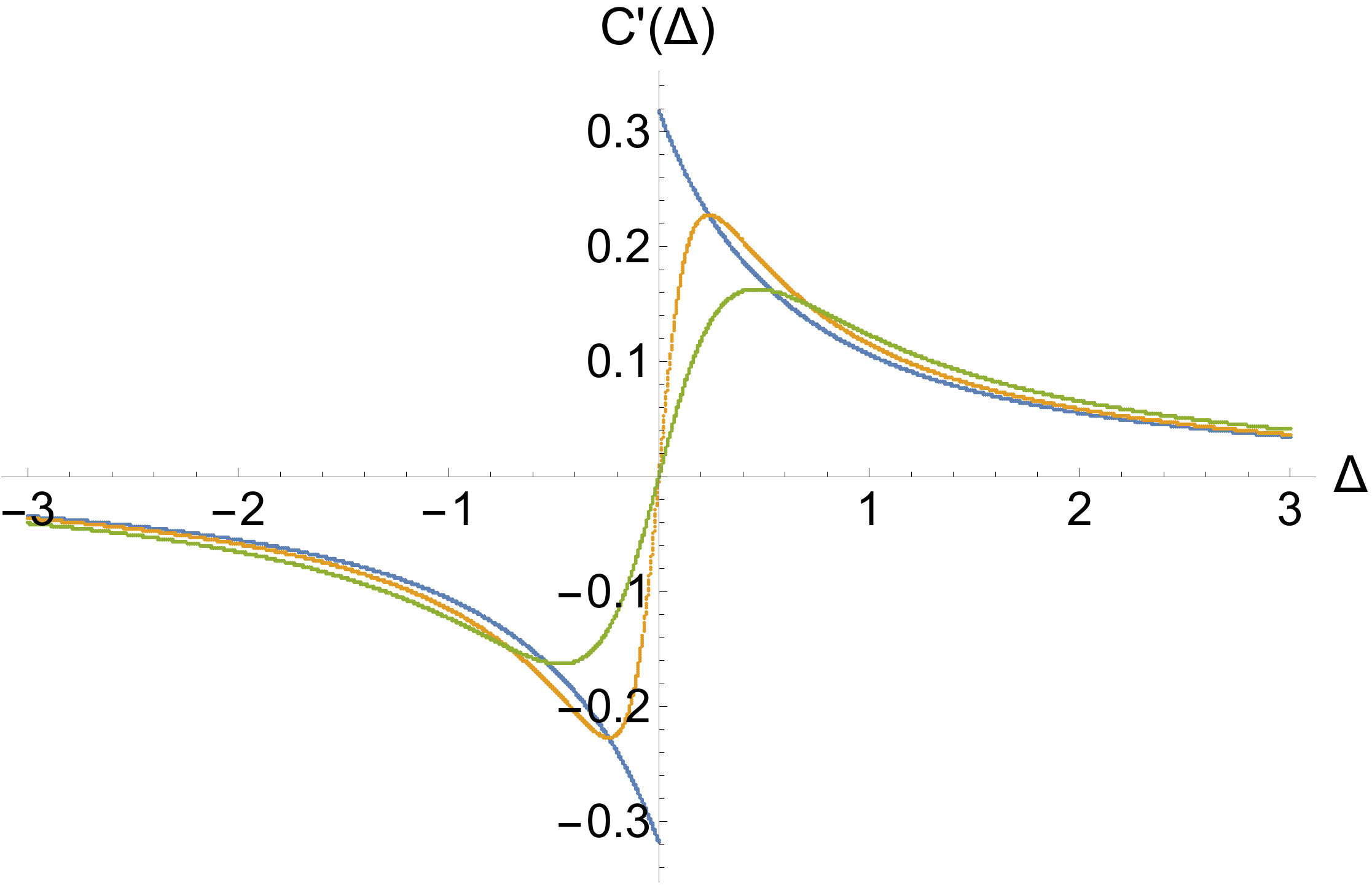}
	\caption{The derivative of spread complexity with respect to $\Delta$ in the continuum limit for the circuit connecting the free fermion ground state and Kitaev chain ground state.  We have set $J=1$.  The blue line has $\mu = 0.98$,  the orange $\mu = 1.02$ and the green $\mu = 1.1$.  When crossing the topological phase transition points at $|\mu| =1$ the derivative develops a discontinuity.}  
	\label{fig:DeltaPlots}
\end{figure}

\noindent
Note that the normalization is with respect  to the total number of sites in the system.  Although we have not found an analytic expression for the continuum integral, it is not difficult to see that
\begin{equation}
C(J, |\mu|< |J|, \Delta) = \frac{1}{2} - \frac{1}{\pi}\int_0^{\frac{\pi}{2} } \sin (\varphi_k) \tan^{-1}\left( \frac{\tan(\varphi_k)}{|\Delta|}   \right)  d\varphi_k\,,
\end{equation} 
which is $\mu$-independent.  In the continuum limit we thus find, qualitatively, the same result as \cite{Caputa:2022eye} where the spread complexity of the circuit connecting the free fermion ground state and Kitaev chain ground state is constant in the topological phase. Consequently, we conclude that spread complexity is indeed sensitive to the phase transition, as can be seen in figure (\ref{fig:Circuit13d}).
\\ \\
To compare with \cite{Caputa:2022eye}, we also plot the spread complexity for a few values of $\Delta$ in Fig. (\ref{fig:InfiniteLCircuit1Deltas}).  The spread complexity takes a constant $\Delta$-dependent value in the topological phase.  At finite $L$ the spread complexity is not constant in the topological phase, but rather resembles a gluing of a discrete number of arc segments as $\mu$ is changed within this phase, as in figure (\ref{fig:FiniteLCircuit1}). As $L$ is increased, the number of these segments increase and the profile approximates the continuum limit constant value that characterizes the topological phase.  
\\ 
Another feature we report on is the behavior of the spread complexity as $\Delta$ approaches zero. From (\ref{CCircuit1}), it is clear that the spread complexity is both continuous and vanishes at $\Delta =0$. The behavior of the first derivative $\frac{\partial C(J, \mu, \Delta )}{\partial \Delta}$, however, is different in the topological and trivial phases as $\Delta$ approaches zero.  Specifically, when $|\mu| > |J|$ the first derivative approaches zero as $\Delta$ approaches zero while for $|\mu| < |J|$ the first derivative approaches a constant value.  Viewed as a function of $\Delta$, running over positive and negative values, the spread complexity develops a cusp in the topological phase at $\Delta = 0$. While we have no physical explanation for this behaviour at present, it is reminiscent of the behaviour of the order parameter of a conventional second order phase transition. In figure (\ref{fig:DeltaPlots}) we  plot the behavior of the derivative for two values of $\mu$ near the phase transition to illustrate this observation.  

\subsection{Circuit 2}

\begin{figure}[h!]
    \centering
    \includegraphics[width=0.7\textwidth]{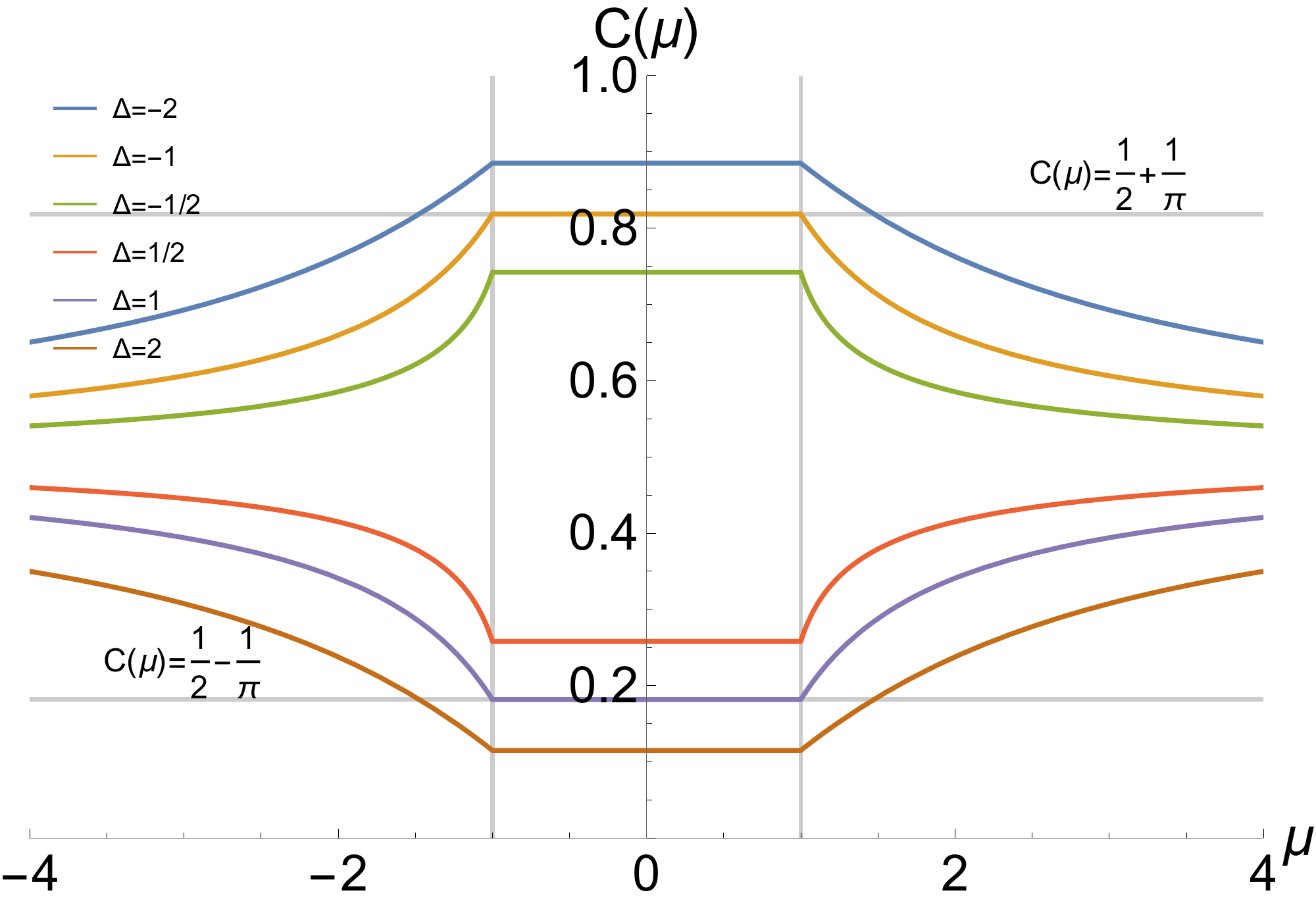}
    \caption{Complexity $\mathcal{C}(s=1; \mu , \Delta)$ as a function of $\mu$ for various choices of $\Delta$ ($\Delta = -2,-1,-1/2,1/2,1,2$) for the circuit connecting the Cooper-pair ground state and Kitaev chain ground state. Between the two vertical gridlines $\mu = \pm 1$, spread complexity is $\mu$-independent. The two dashed horizontal gridlines are the analytical results of $\mathcal{C}(|\mu|<1, \Delta = \pm 1)$, respectively. For $|\mu| \to \infty$, the spread complexities of various $\Delta$s approach the dotted horizontal gridline $\mathcal{C} =1/2$.}
    \label{fig:Kt_cmpl_mu}
\end{figure}

\begin{figure}
    \centering
    \includegraphics[width=0.7\textwidth]{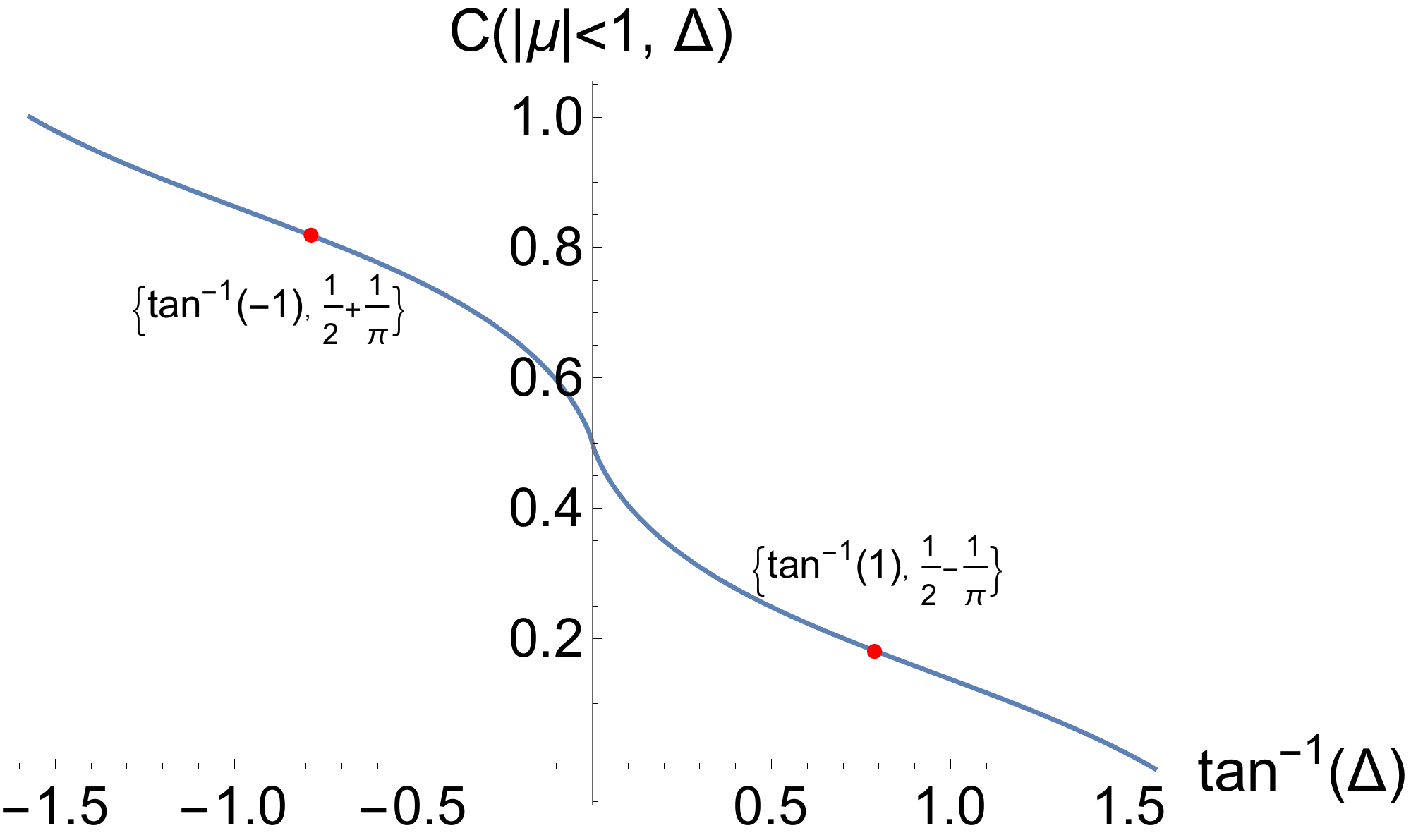}
    \caption{Complexity $\mathcal{C}(s=1; |\mu|<1 , \Delta)$ as a function of $\Delta$ for the circuit connecting the Cooper-pair ground state and Kitaev chain ground state. The two red dots are the analytical results of $\mathcal{C}(|\mu|<1, \Delta = \pm 1)$, respectively.}
    \label{fig:Kt_cmpl_Del}
\end{figure}

The spread complexity of the circuit in section \ref{pkccHm} exhibits a plateau-like feature when the system is in the topological phase. This is reminiscent of results obtained for the SSH model \cite{Caputa:2022eye}, where the spread complexity also plateaued in the topological phase.  An undesirable feature of the first circuit, however, is that the reference state (\ref{gsM}) is $\mu$-dependent.  A natural question to ask is whether we can make a $(\mu,\Delta)$-independent choice for the reference state such that the spread complexity exhibits a plateau in topological phase? The answer it turns out is yes, if we select reference state to be
\begin{equation}
|\Omega_k (s=0) \rangle  = \exp \left[ \frac{\pi}{4} e^{-i \frac{\pi}{2}} J_+^{(k)} - \frac{\pi}{4} e^{+i \frac{\pi}{2}} J_-^{(k)} \right] \left|\frac{1}{2}, -\frac{1}{2} \right\rangle_k.
\end{equation}
This state can be interpreted physically as the ground state of the Cooper-pair (fermion interaction) part of the Kitaev Hamiltonian. With this choice of reference state the circuit takes the form
\begin{equation}
|\Omega_k(s)\rangle =   \exp \left[ -i \cdot s \cdot  \left( \operatorname{sgn} \Delta \cdot \frac{\pi-\varphi_k}{2} -\frac{\pi}{4} \right) \left( J_+^{(k)} + J_-^{(k)} \right)  \right] |\Omega_k (s=0) \rangle\,,
\end{equation}
from which we compute the spread complexity to be
\begin{equation}
\mathcal{C}_k(s=1) = \sin^2 \left( \operatorname{sgn} \Delta \cdot \frac{\pi-\varphi_k}{2} -\frac{\pi}{4} \right) = \frac{1}{2}\left(1 - \frac{\Delta \sin k}{\sqrt{(\mu + J \cos(k) )^2 + \Delta^2 \sin^2(k)  }}\right)\,.   
\end{equation}
Integrating by parts, we find that in the regime $|\mu |< |J|$ the total spread complexity is given by
\begin{equation}
\mathcal{C} (J, |\mu|<|J| , \Delta) 
= \frac{1 - \operatorname{sgn} \Delta}{2} + \frac{1}{\pi} \int_0^{\pi/2}\,d\varphi_k\, \cos \varphi_k\, \tan^{-1} \frac{\tan \varphi_k}{\Delta}\,. 
\end{equation}
In the topological phase the spread complexity is therefore independent of the chemical potential, $\mu$.  The integral can be performed analytically to give
\begin{equation}
\mathcal{C} (J, |\mu|<|J| , \Delta)
= \frac{1}{2} - \frac{1}{\pi} \frac{\Delta \tan^{-1} \sqrt{\Delta^2-1}}{ \sqrt{\Delta^2 -1}}\,. \label{plateauVal}
\end{equation}
We plot the spread complexity as a function of $\mu$ for various choices of $\Delta$ in Fig. (\ref{fig:Kt_cmpl_mu}).  The plateau feature of spread complexity in the topological phase can be clearly observed.  In fig. (\ref{fig:Kt_cmpl_Del}) we plot the dependence on $\Delta$ of spread complexity inside the topological phase i.e. the height of the plateau.  The red dots on the figure coincide with the plateau value for the SSH model \cite{Caputa:2022eye}, at $\Delta = - 1$.  At these points the numerator and denominator of (\ref{plateauVal}) tend to zero, so that the value is obtained by taking the limit $\Delta^2 \rightarrow 1$. \\ \\
As in the previous case, the derivatives of spread complexity, in the topological phase, also exhibit noteworthy behaviour.  Specifically, the derivative of complexity with respect to $\Delta$,
\begin{equation}
\pi \frac{d}{d\Delta} \mathcal{C} (J, |\mu|<|J| , \Delta) = \frac{1}{1-\Delta^2}+\frac{\tan ^{-1}\left(\sqrt{\Delta^2-1}\right)}{\left(\Delta^2-1\right)^{3/2}},
\end{equation}
is continuous everywhere, including $|\Delta|=1$. The second derivative is also continuous at $|\Delta|=1$ but diverges at $\Delta =0$, since
\begin{equation}
\pi \frac{d^2}{d\Delta^2} \mathcal{C} (J, |\mu|<|J| , \Delta) =\frac{2 \Delta ^2+1}{\Delta  \left(\Delta ^2-1\right)^2}-\frac{3 \Delta  \tan ^{-1}\left(\sqrt{\Delta ^2-1}\right)}{\left(\Delta ^2-1\right)^{5/2}}\sim \Delta^{-1}\,,
\end{equation}
as $\Delta \to 0$. Again, note that this curious divergence of the second derivative is reminiscent of a third order (conventional) phase  transition and deserves further investigation that we leave for future explorations.

\subsection{Circuit 3}\label{pkccV}

\begin{figure}[!h]
\begin{subfigure}{.5\textwidth}
	\centering
		\includegraphics[width=0.90\textwidth]{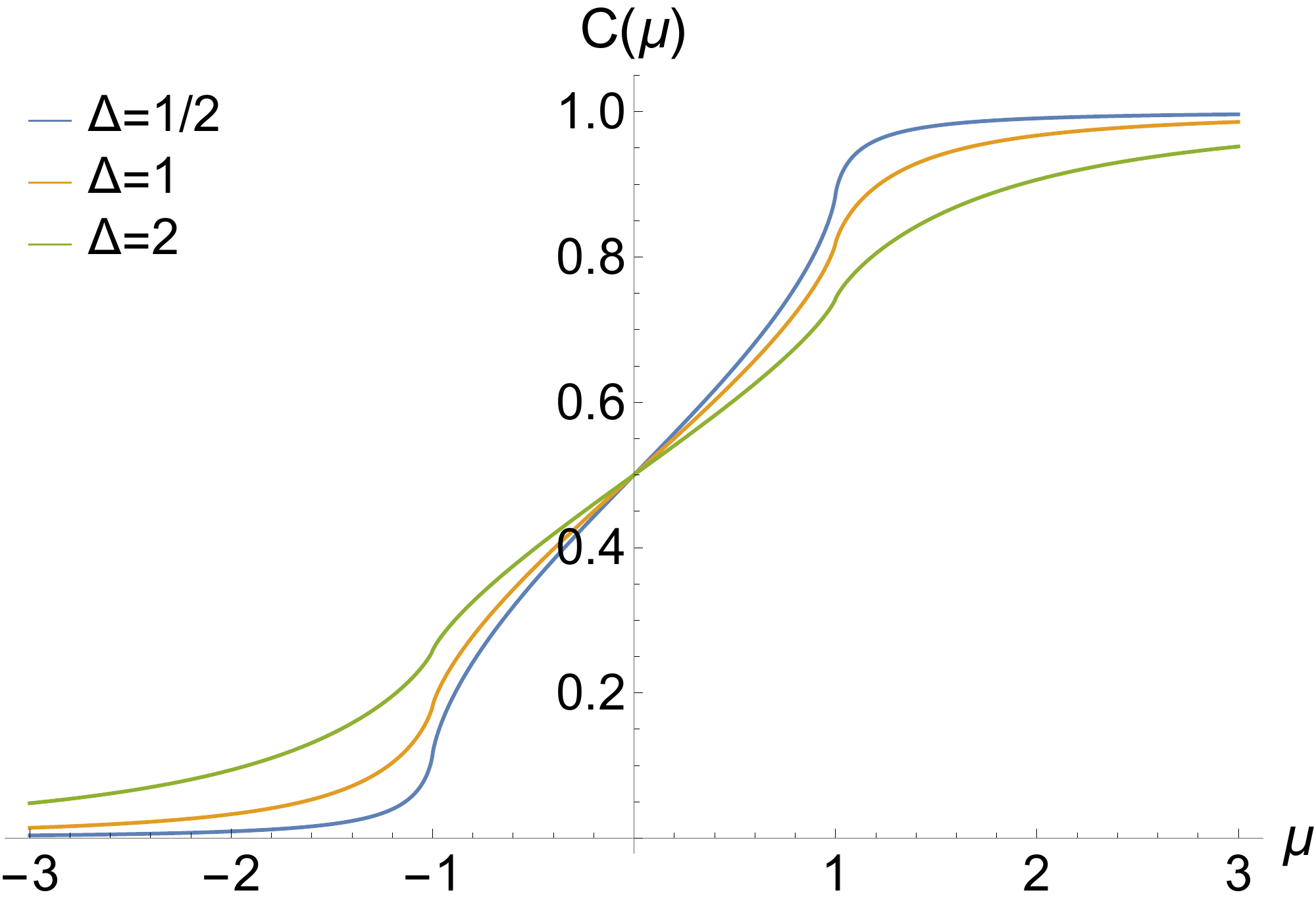}
	\caption{ }
	\label{fig:spreadComplexityCircuit2}
	\end{subfigure}
	\begin{subfigure}{.5\textwidth}
		\centering
		\includegraphics[width=0.90\textwidth]{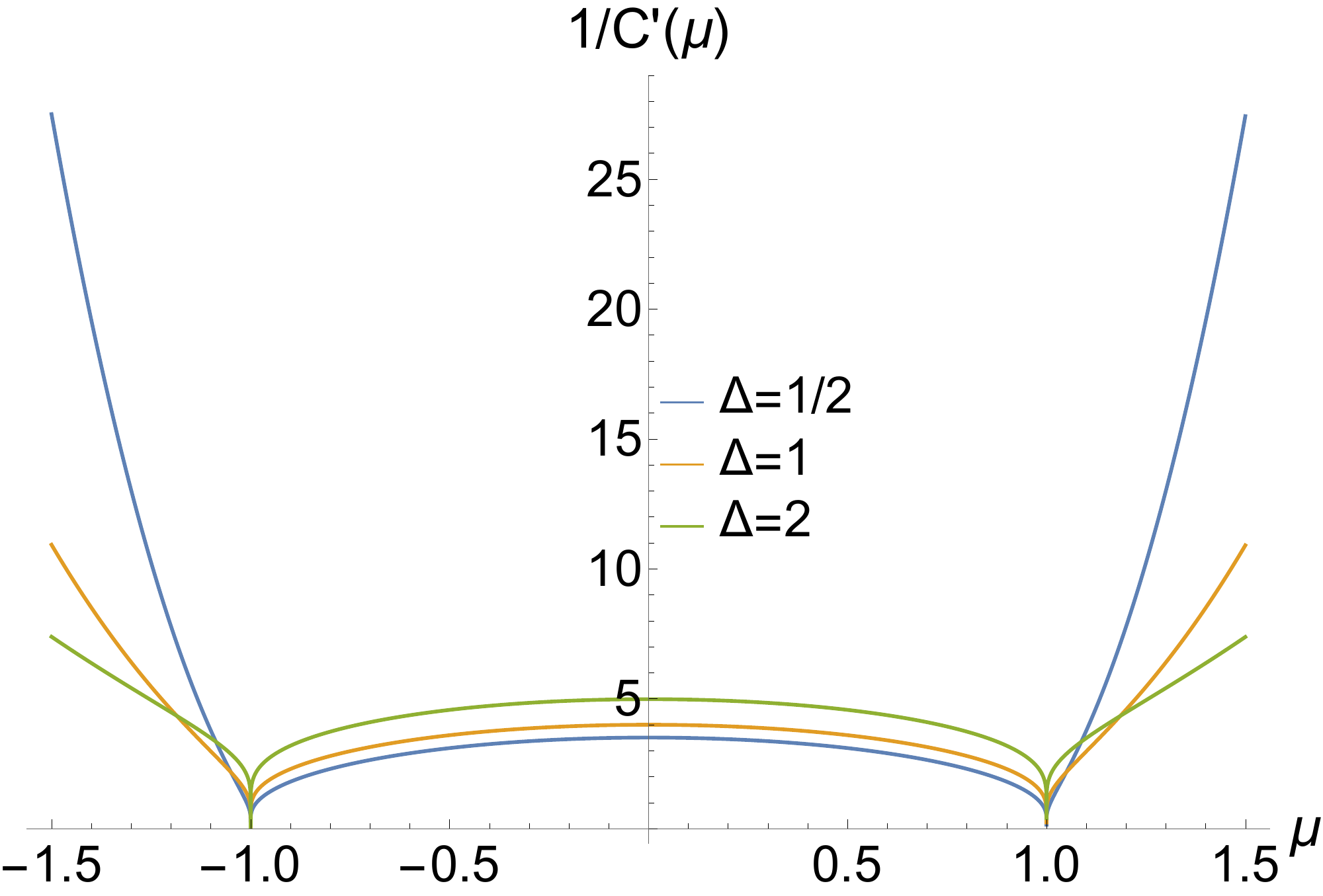}
	\caption{ }
	\label{fig:DerivativeComplexity}
	\end{subfigure}
	\caption{a) The spread complexity for the circuit connecting the vacuum and the Kitaev chain ground state in the continuum limit.  We have set $J=1$.  The spread complexity is monotonically increasing as a function of $\mu$.  b) The inverse of the derivative of the spread complexity as a function of $\mu$ for $\Delta=\frac{1}{2}$ and $J=1$.  At the phase transition the derivative diverges so that its inverse approaches zero.}
\end{figure}

Finally, let's consider the circuit connecting the fermion vacuum to the Kitaev chain ground state. The fermion vacuum, with its complete absence of particle states, is in some sense the most straightforward choice of reference state. In the $SU(2)$ representation it is the tensor product (over momenta) of lowest weight spin-$\frac{1}{2}$ states. For each value of $k$ the circuit takes the form
\begin{equation}
|\Omega_k(s)\rangle =   \exp \left[ s \cdot \left( \frac{\pi-\varphi_k}{2} e^{-i \left(\frac{\pi}{2} -\operatorname{arg} \Delta \right)} J_+^{(k)} -\frac{\pi-\varphi_k}{2} e^{i \left(\frac{\pi}{2} -\operatorname{arg} \Delta \right)} J_-^{(k)} \right) \right] \left|\frac{1}{2}, -\frac{1}{2} \right\rangle_k,
\end{equation}
where we have again made use of  Baker-Campbell-Hausdorff identities to rewrite the ground state \eqref{grdstt}. We obtain the spread complexity as
\begin{equation}
\mathcal{C}_k(s=1) = \sin^2 \frac{\pi - \varphi_k}{2} = \frac{1}{2}\left(1 + \frac{\mu + J \cos(k)}{\sqrt{(\mu + J \cos(k) )^2 + \Delta^2 \sin^2(k)  }}\right) . 
\label{CkCircuit2}
\end{equation}
The derivative of the spread complexity with respect to $\mu$,
\begin{equation}
\frac{\partial}{\partial \mu} \mathcal{C}_k(s=1) = \frac{ \Delta^2\sin^2(k)}{2(  (\mu + J \cos(k))^2 +   \Delta^2 \sin^2(k)  )^{\frac{3}{2}} }\,,
\end{equation}
will also prove important.
As before, we need to sum over the values of $k$ in order to obtain the spread complexity for this circuit. In figure (\ref{fig:spreadComplexityCircuit2}), we plot the spread complexity as a function of $\mu$ in the continuum limit. Surprisingly, unlike for the previous circuits, the complexity monotonically increases with $\mu$. In particular we do not observe any plateau in the spread complexity in the topological phase. The spread complexity still has a noteworthy feature at the phase transition point.  Specifically, the first derivative of complexity diverges at the phase transition points, as in figure (\ref{fig:DerivativeComplexity}). The divergence at $\mu = -1$ is due to low momentum contributions while the divergence at $\mu=1$ is due to the high momentum contributions, 
\begin{eqnarray}
    \left. \frac{\partial }{\partial \mu} \mathcal{C}_k(s=1) \right|_{\mu = 1} & = & \frac{1}{2 |\Delta| k} + \mathcal{O}(k)\,,   \nonumber  \\
   \left. \frac{\partial }{\partial \mu} \mathcal{C}_k(s=1) \right|_{\mu = -1} & = & \frac{1}{2 |\Delta| (k - \pi)} + \mathcal{O}(k - \pi)\,.  
\end{eqnarray}
For the intermediate regime $|\mu| <1$ there always exists a pole at some value of $k \in \left[0, \pi \right]$ at which the integral diverges. Evidently, even though we do not observe a plataeu in the topological phase, the spread complexity for this circuit is able to distinguish the topological phase transition in the chain.

\noindent
At finite $L$ we observe that the spread complexity has the same schematic form depicted in (\ref{fig:spreadComplexityCircuit2}).  The derivative, however, remains finite, with a sharp change in the sign of the second derivative.  As $L$ increases the derivative increases, reaching the diverging value in the $L \rightarrow \infty$ limit. In this circuit the behavior around $\Delta = 0$ is also an alternate diagnostic of the topological phase transition.  Using (\ref{CkCircuit2}) we may compute 
\begin{eqnarray}
    \left.  \frac{\partial^2 }{\partial \Delta^2} C_k(s=1)  \right|_{\mu = -1, \Delta =0} & = & \frac{1}{2} \cot^2\left( \frac{k}{2}\right) ,  \nonumber  \\
        \left.  \frac{\partial^2 }{\partial \Delta^2} C_k(s=1)  \right|_{\mu = -1, \Delta =0} & = & -\frac{1}{2} \tan^2\left( \frac{k}{2}\right) ,
\end{eqnarray}
which diverges when integrated over the interval $k \in \left[0, \pi\right]$.  In the region $|\mu| > 1$ we find that the second derivative remains finite while in the region $|\mu| \leq 1$ the integral diverges.  As in the first circuit, a pole develops in the second derivative at some value of $k$ inside the integration range, leading to a diverging integral.   

\subsection{Odd $L$ and Periodic boundary conditions}
The results in the previous three subsections were obtained by studying the APBC Kitaev chain with an even number of sites and afterwards taking the limit to an infinite number of sites.  One may wonder how this choice of boundary condition and the parity of sites affect the results.  In this subsection we  briefly explain why the large $L$ limit results are insensitive to these choices. \\ \\
In the case of odd $L$ and/or a periodic boundary condition on the Kitaev chain, the momenta (\ref{kVals}) can include $k = 0$ and/or $k = \pi$.  Since, for these values, $a_k = a_{-k}$, this situation requires special attention. This adds one of or both of the terms       \begin{eqnarray}
    H_{0} & = & -(\mu + J) (a_0^\dag a_0 - a_0 a_0^\dag),   \nonumber \\
    H_{\pi} & = & -(\mu - J) (a_\pi^\dag a_\pi - a_\pi a_\pi^\dag),
\end{eqnarray}  
to the Hamiltonian (\ref{Hsu2Basis}), see e.g. Appendix A of \cite{Kawabata:2017zsb}.  The eigenstates of $H_0, H_{\pi}$ is simply the number operator basis i.e. a single fermion or no fermion.  The ground state for $H_0$ and $H_{\pi}$ is given by 
\begin{eqnarray}
    |\Omega\rangle_0 & = & |0\rangle_0 + \Theta( \mu + J  )\left(a_0^\dag|0\rangle_0 - |0\rangle_0 \right),  \nonumber \\
     |\Omega\rangle_\pi & = & |0\rangle_\pi + \Theta( \mu - J  )\left(a_\pi^\dag|0\rangle_\pi - |0\rangle_\pi \right),
\end{eqnarray}
respectively.  The spread complexity can be computed using the Heisenberg algebra formulae of \cite{Caputa:2021sib} which adds a factor of $1$ to the total complexity if the ground state differs from the reference state by a single fermion.\\

\noindent
In the large $L$ limit this contribution is subleading and the results of the previous subsections hold without modification.  At finite $L$ we need to add $1$ to the complexity at $\mu = -J$ ($\mu = J$) if $H_0$ ($H_{\pi}$) is included for the boundary condition under consideration. This shift has the effect of transforming the finite-$L$ spread complexity from a continuous function to a discontinuous one at these points.


\section{Discussion}
\label{sec:dis}
Computational complexity, a well-known entity in the context of computer science and associated mathematical subfields like graph theory, is still very much a shiny new concept in high energy physics. While the primary motivation for its introduction in the latter was the understanding of the black hole information-loss paradox, it has proven far more versatile with recent applications to the emergence of spacetime in quantum gravity, early universe cosmology \cite{Bhattacharyya:2020rpy,Bhattacharyya:2020kgu,Haque:2021kdm,Haque:2021hyw} and even condensed matter physics. It is this last aspect that is the focus of this article.\\

\noindent
Specifically, this paper details our study of a particular variant of quantum complexity called spread complexity, introduced in \cite{Balasubramanian:2022tpr}, applied to Kitaev's 1-dimensional p-wave superconductor. The Kitaev chain is a prototypical example of a system exhibiting topological phase transitions, and an important laboratory for tests of novel features of quantum systems such as topological order and quantum phases of matter. Building on the methodology of \cite{Caputa:2022eye}, we show here that spread complexity, based on the associated Krylov basis of the Kitaev chain, is sensitive to the topological phase transition exhibited by the model. Our results are in concordance with, and provides further support for, those of \cite{Caputa:2022eye}, where results for the SSH model were also in the affirmative for spread complexity as a novel probe of the topological phase transition. \\ \\ 
Furthermore we have demonstrated a level of robustness of these results by computing the spread complexity for circuits of differing reference states connected to the Kitaev ground state taken as the fixed target state.  When using the $\Delta=0$ or $J = 0$ ground state as reference state we have shown that, in the topological phase, the spread complexity is a constant as a function of $\mu$.  Additionally, its second derivative with respect to $\Delta$ exhibits divergences when the system is in the topological phase.  When using the vacuum as reference state the spread complexity does not exhibit the plateau feature, but its first derivative with respect to $\mu$ diverges at the  transition point.  In all these cases the spread complexity thus does, in some way, capture the transition from the trivial to topological phase.  In essence, the statement that we are making is that two states realizing distinct topological phases, certainly in the Kitaev chain and the SSH models cannot be mapped to one another via low-depth circuits.
\\ \\
The circuits that we have considered connect simple choices of reference state to the ground state of the Kitaev chain.  Even though the spread complexity does indeed capture the transition from trivial phase to topological phase in some way for each of our circuit choices, it is clear that not all choices exhibit a plateau in the topological phase.  Based on results for the SSH model \cite{Caputa:2022eye} this may seem surprising, and it leads to a natural question: which circuit choices give rise to a spread complexity plateau?  Presumably, an understanding of this feature should be related to the symmetries of states in different phases and it may be possible to classify circuits according to whether they preserve or break these symmetries. Hopefully, elucidating this symmetry classification may provide some hint as to the general choices of circuits that would display the plateau feature in systems exhibiting topological phase transitions. 
\\ \\
The utility of spread complexity as an efficient probe of quantum phase transitions can be extended and explored in several different directions:
\begin{itemize}
\item
Given the divergence of fluctuations at a quantum critical point, it might intuitively be expected that the complexity of say the ground state of a system undergoing a quantum phase transition diverges at the transition. This article reflects our attempts to sharpen this intuition and, like its predecessors in the literature \cite{Ali:2018aon,Liu:2019aji,Caputa:2021sib} focused on a particular type of quantum phase transition in which the transition is between states realizing distinct topological phases. Here we can indeed make a crisp statement about the complexity (and its derivatives) at the transition. However, topological phase transitions are just one in an ever-expanding list of quantum phase transitions that include conventional (Landau) order, Berezinskii-Kosterlitz-Thouless-transitions 
\cite{Kosterlitz_1973} and the more exotic deconfined critical points of \cite{Senthil_2004}. It would be extremely exciting if spread complexity, or the closely related K-complexity offered a window into these phase transitions as well. Indeed one tantalising possibility, which we leave for future work, might lie in the well-known relation between the Kitaev chain and the transverse field Ising model that map the topological order of the former to conventional order in the latter.
\item
The computations in this article rely strongly on us being able to represent the Kitaev Hamiltonian in a basis of $su(2)$ generators. But this is not always possible; a fact that is demonstrable even in the Kitaev chain itself. For example, when the (anti)periodic boundary conditions studied above are extended to more general {\it twisted} boundary conditions, the Hamiltonian can no longer be written in terms of $su(2)$ generators. Understanding how to compute the spread complexity in such situations would be an important development of this computational framework. As a first step in this direction, in Appendix B, as a preview of forthcoming work, we give some preliminary results for the Kitaev chain with twisted boundary conditions.
\end{itemize}

\acknowledgments

JM would like to acknowledge support from the ICTP through the Associates Programme and from the Simons Foundation through grant number 284558FY19 as well as funding from the South African Research Chairs
Initiative funded by the National Research Foundation and Department of Science and Technology. NG is supported by a Faculty of Science PhD Fellowship. H.J.R.vZ is supported by the ``Quantum Technologies for Sustainable Devlopment" grant from the National Institute for Theoretical and Computational Sciences (NITHECS). Work of PC and SL is supported by “Polish Returns 2019” grant of the National Agency for Academic Exchange (NAWA) PPN/PPO/2019/1/00010/U/0001 and Sonata Bis 9 2019/34/E/ST2/00123 grant from the National Science Centre, NCN.

\bibliography{kitaevchainrefs.bib}

\appendix

\section{$\pmb{SU(2)}$ coherent states}
\label{su2Appendix}

Following \cite{Caputa:2021sib} we present here an explicit identification of the K-complexity for $SU(2)$ coherent states with an $SU(2)$ operator expectation value.  Similar results may be obtained for $SU(1,1)$ and Glauber coherent states.   Given a Liouvillian
\begin{equation}
L = \alpha J_{+} + \alpha^* J_{-} + \gamma J_0, 
\end{equation}
time-evolving the $SU(2)$ lowest weight state we have
\begin{equation}
e^{i L t} |j,-j\rangle = \mathcal{N} e^{z(\alpha, \alpha^*, \gamma) J_{+}} |j,-j\rangle \equiv \frac{1}{\sqrt{(\bar{z} | z) } }   |z),
\end{equation}
by using, for example, Baker-Campbell-Haussdorff formulae. The number $\mathcal{N}$ is the normalisation.  The use of coherent states make the Krylov basis immediately apparent as
\begin{equation}
|O_n\rangle  = \frac{1}{\sqrt{\langle j,-j | J_{-}^n J_{+}^n |j,-j\rangle } } J_{+}^n |j,-j\rangle ,
\end{equation}
so that the Krylov complexity operator is 
\begin{equation}
\hat{K}_n =  \sum_{n=0} n \frac{J_{+}^n |j,-j\rangle \langle j,-j | J_{-}^n }{\langle j,-j | J_{-}^n J_{+}^n |j,-j \rangle }.  
\end{equation} 
For the $SU(2)$ coherent state the Krylov complexity is thus 
\begin{eqnarray}
C & = & \frac{1}{(\bar{z}|z )}  (\bar{z}| \hat{K}_n |z),   \nonumber \\
  & = & \sum_{n,m,m'} \frac{n}{(m!)(m'!)} \frac{\langle \psi_0 |  \bar{z}^m L_{-}^m    L_{+}^n |\psi_0\rangle \langle \psi_0 | L_{-}^n z^{m'} L_{+}^{m'}       |  \psi_0 \rangle}{(\bar{z}| z)  \langle \psi_0 | L_{-}^n L_{+}^n   | \psi_0 \rangle } ,  \nonumber \\
	& = & \sum_{n } \frac{n}{(n!)^2} \bar{z}^n z^{n}   \frac{ \langle \psi_0 | L_{-}^n L_{+}^n   | \psi_0 \rangle }{(\bar{z}| z)},  \nonumber \\
	& = & z \partial_z \log (\bar{z} | z ),
\end{eqnarray}
which is exactly equal to the normalised expectation value
\begin{equation}
C = \langle \bar{z} | J_0 + j |z\rangle.
\end{equation}

\section{Twisted boundary Kitaev chain}
\label{twisted-appendix}

\begin{figure}[!h]
\begin{subfigure}{.49\textwidth}
	\centering
		\includegraphics[width=0.90\textwidth]{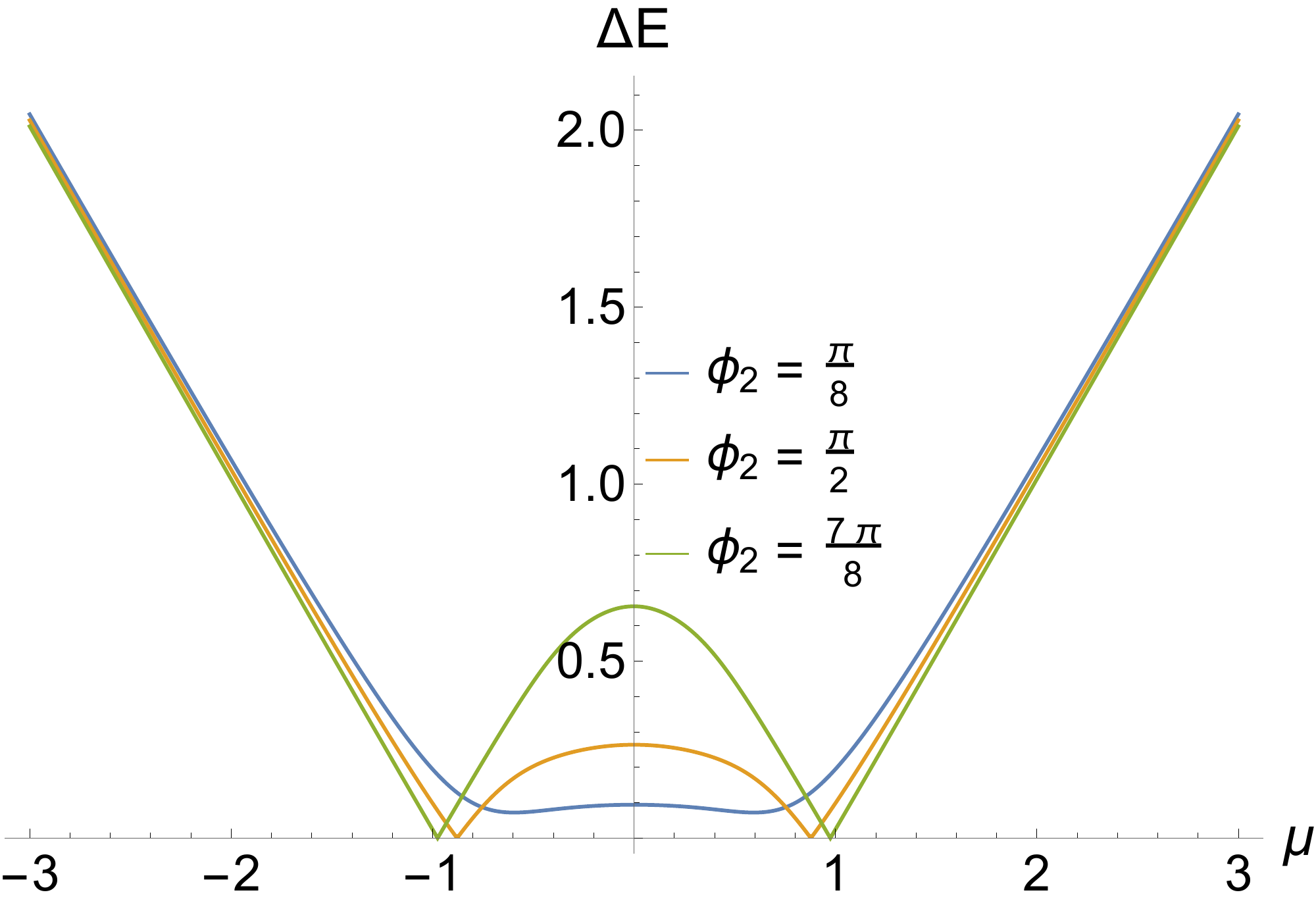}
	\caption{ }
	\label{fig:spreadComplexityTwist}
	\end{subfigure}
	\begin{subfigure}{.49\textwidth}
		\centering
		\includegraphics[width=0.90\textwidth]{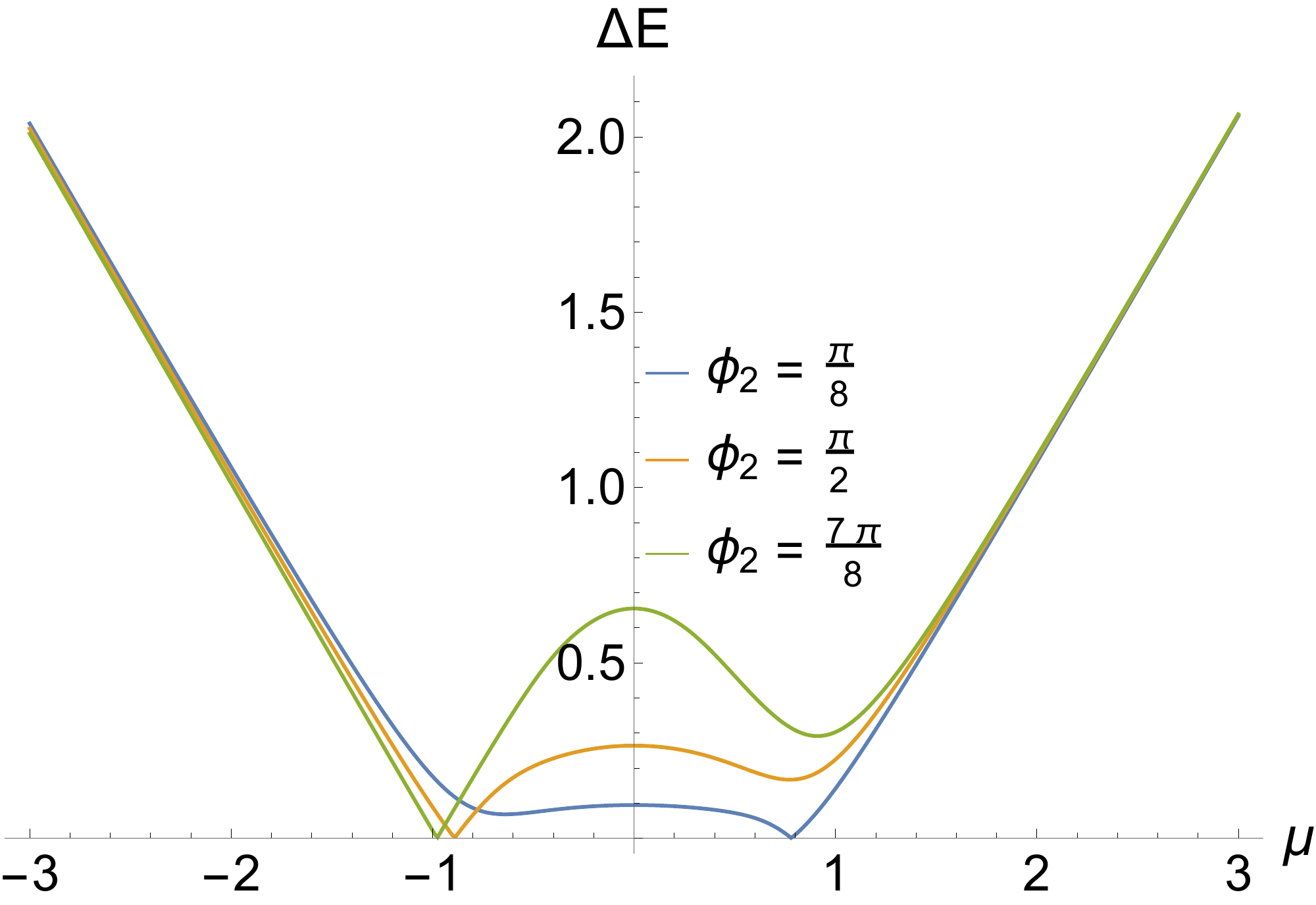}
	\caption{ }
	\label{fig:EnergyGap}
	\end{subfigure}
	\caption{The energy gap between the ground state and first excited as a function of $\mu$ with $\Delta=J=1$, $a=b=1$.  We have fixed $\phi_1 = -\frac{ \pi}{4}$.   a) The chain length is $L=8$.  For these parameters we note a pair of values for $\mu$ exhibit a closing energy gap provided $|\phi_2| > |\phi_1|$    b) The chain length is $L=9$. In this case we observe only a single value for $\mu$ for which gap closing occurs.}
\end{figure}

\begin{figure}
\begin{minipage}{0.32 \textwidth}
    \centering
    \subfloat[$L=8, \phi_2 = \frac{\pi}{8}$]{\includegraphics[width=0.9\textwidth]{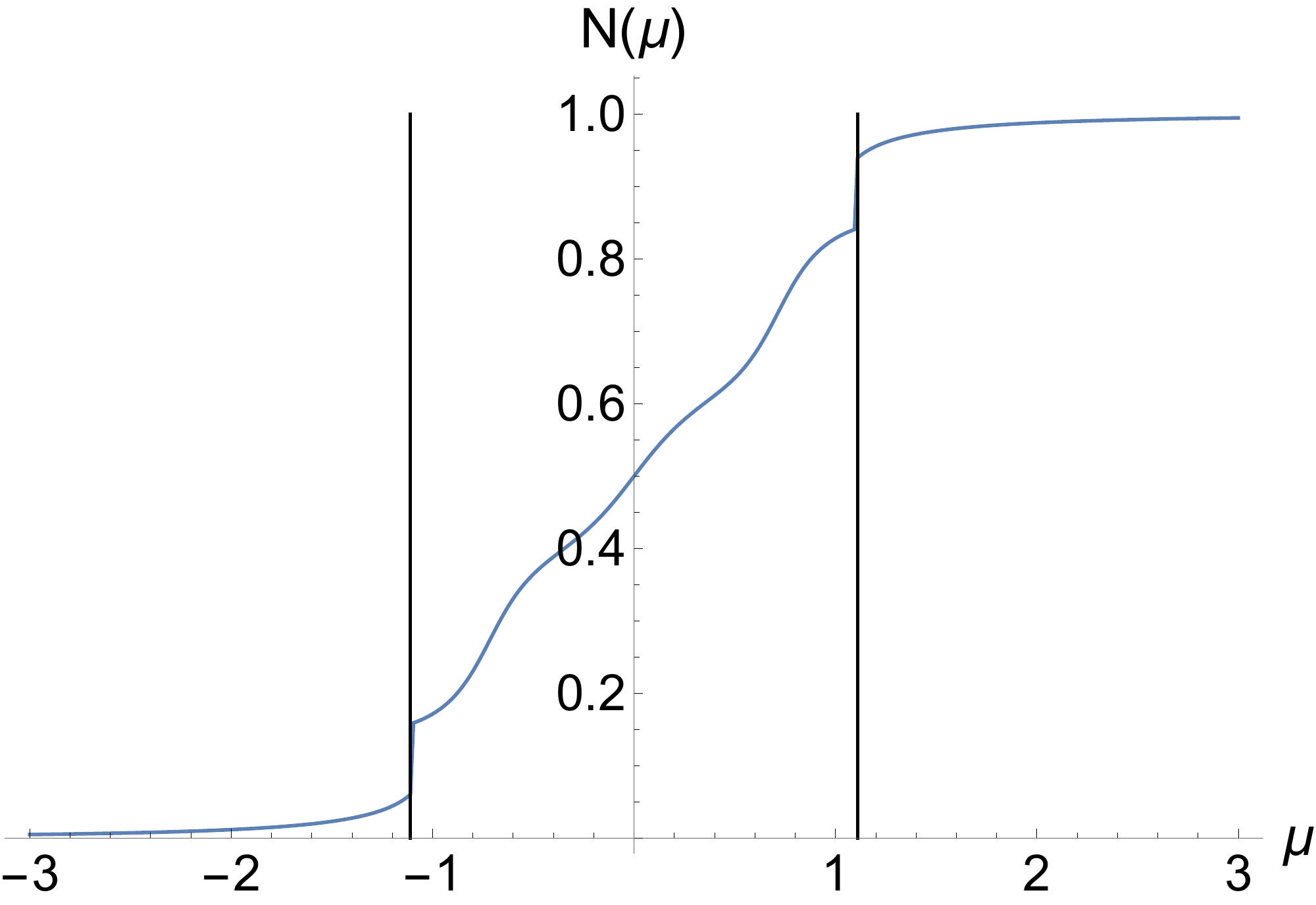} }
    \vfill
     \subfloat[$L= 7, \phi_2 = \frac{\pi}{8}$]{\includegraphics[width=0.9\textwidth]{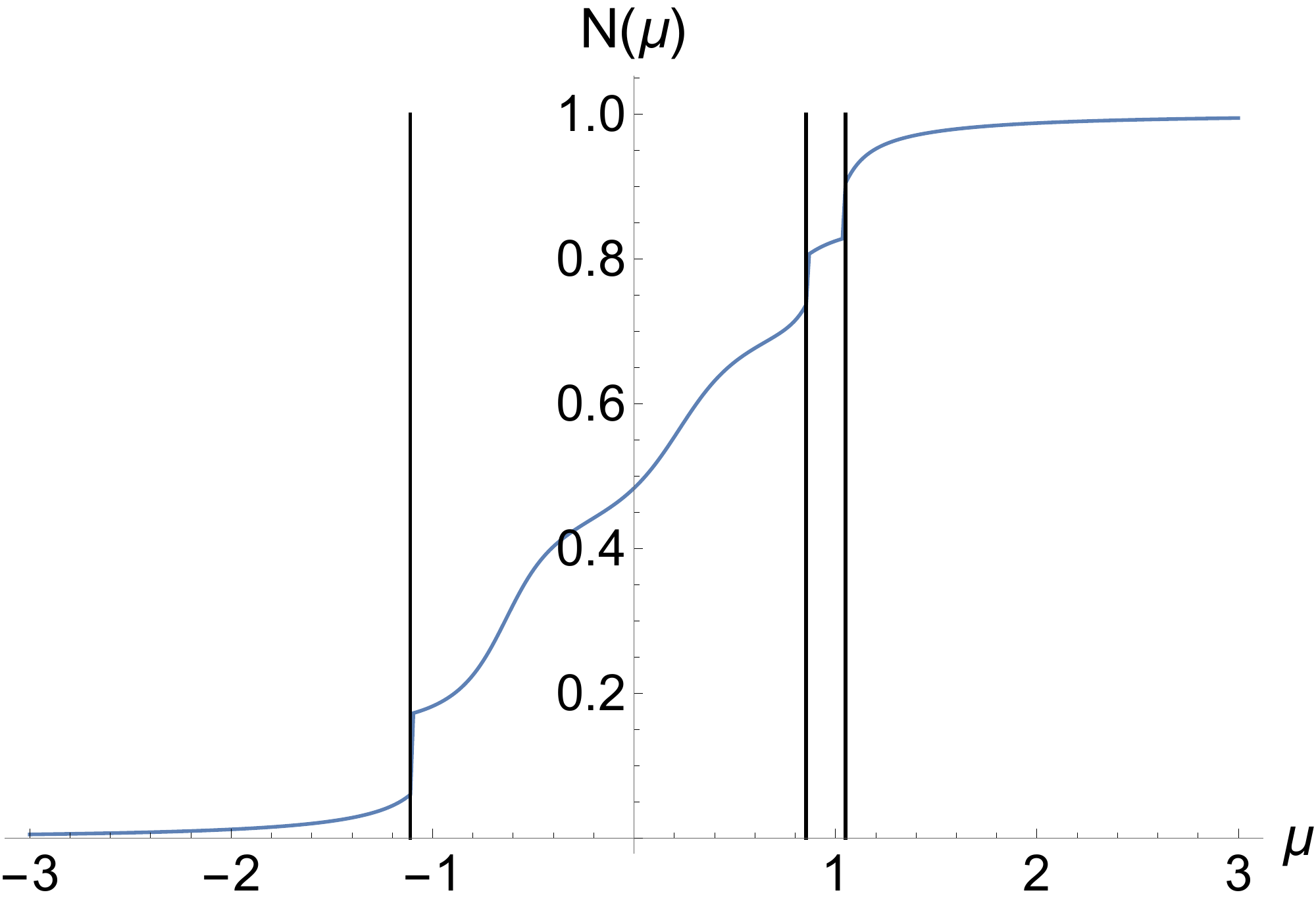} }
\end{minipage}
\begin{minipage}{0.32 \textwidth}
    \centering
    \subfloat[$L= 8, \phi_2 = \frac{\pi}{2}$]{\includegraphics[width=0.9\textwidth]{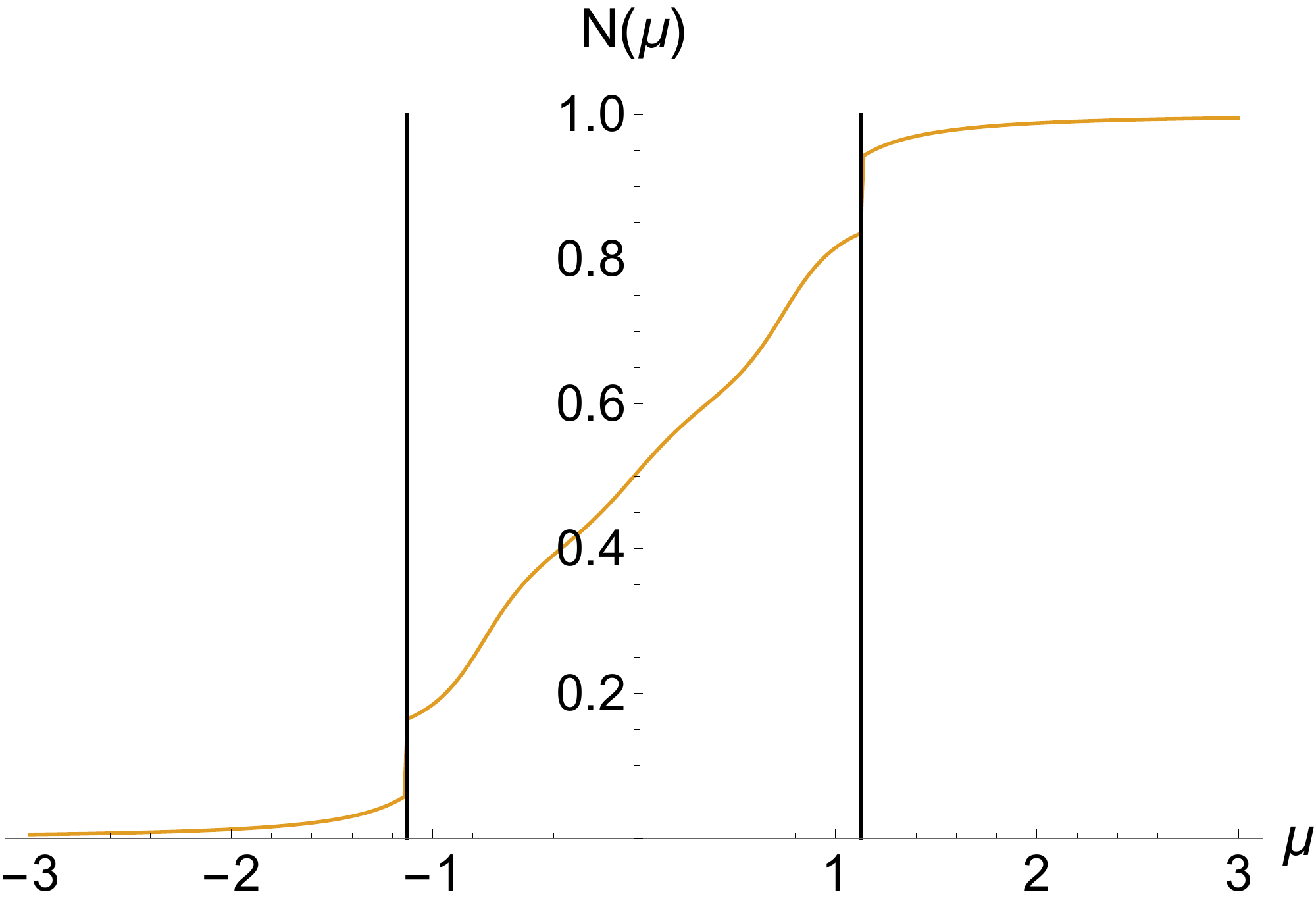} }
    \vfill
     \subfloat[$L= 7, \phi_2 = \frac{\pi}{2}$]{\includegraphics[width=0.9\textwidth]{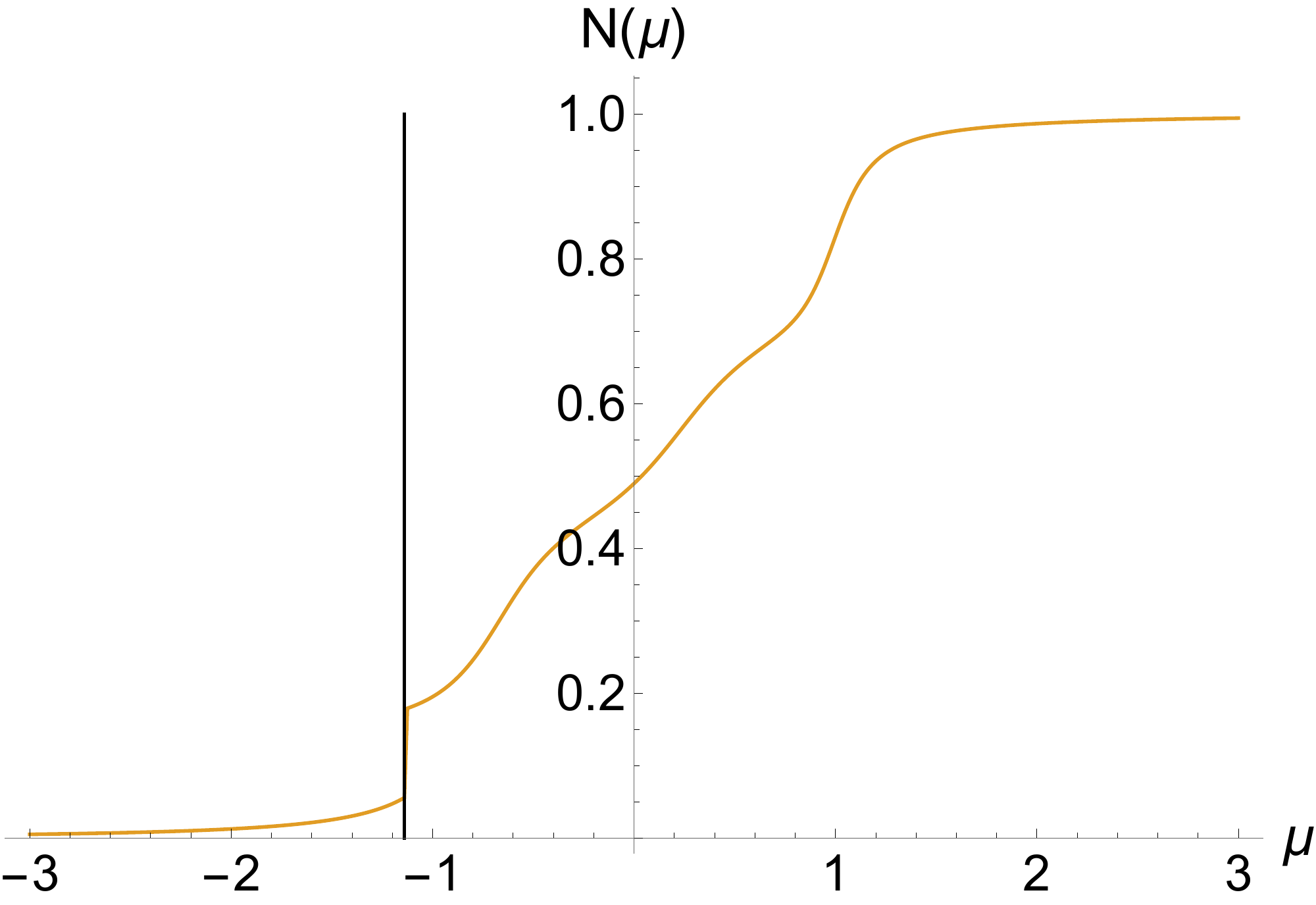} }
\end{minipage}
\begin{minipage}{0.32 \textwidth}
    \centering
    \subfloat[$L=8, \phi_2 = \frac{7\pi}{8}$]{\includegraphics[width=0.9\textwidth]{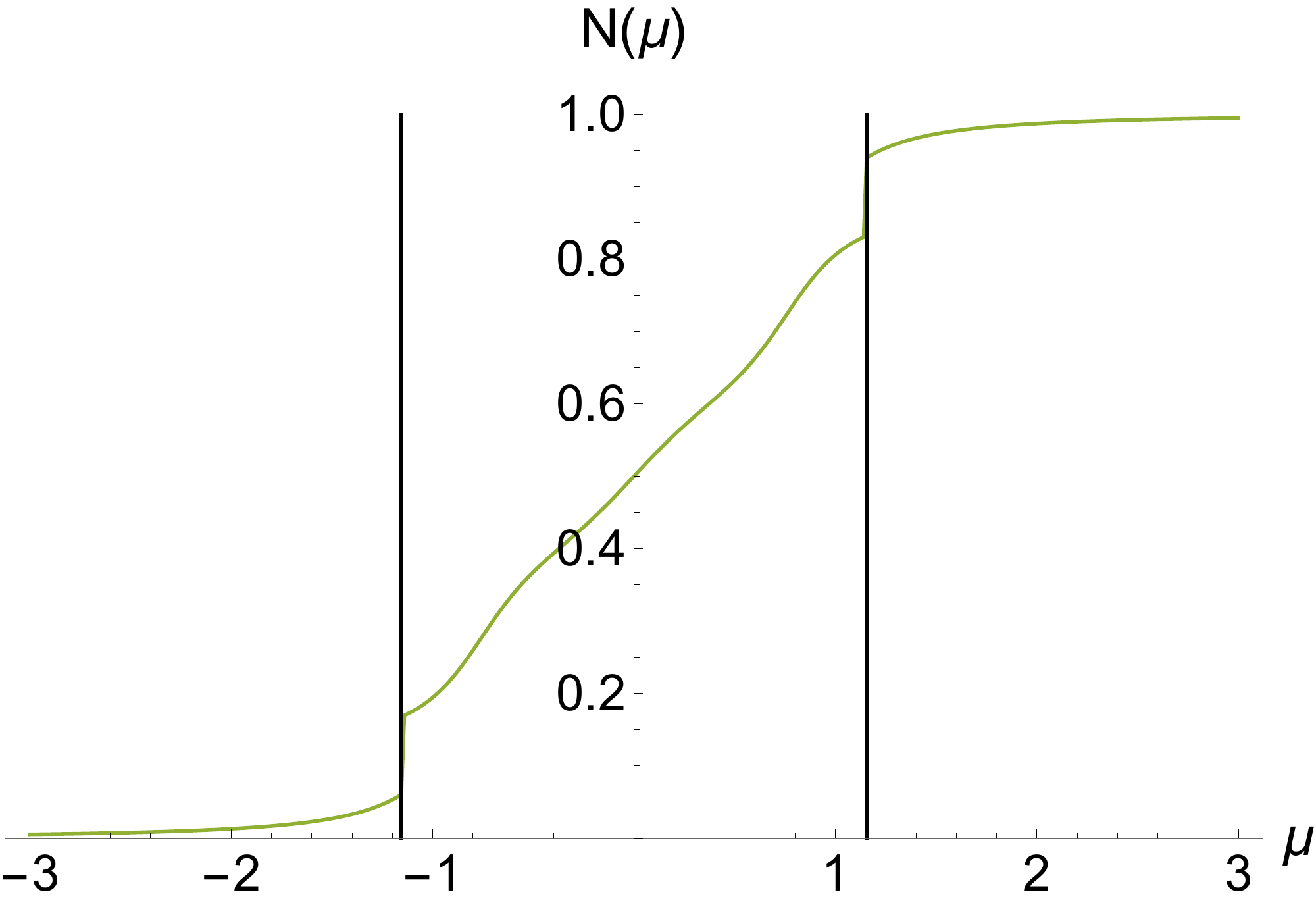} }
    \vfill
     \subfloat[$L= 7, \phi_2 = \frac{7\pi}{8}$]{\includegraphics[width=0.9\textwidth]{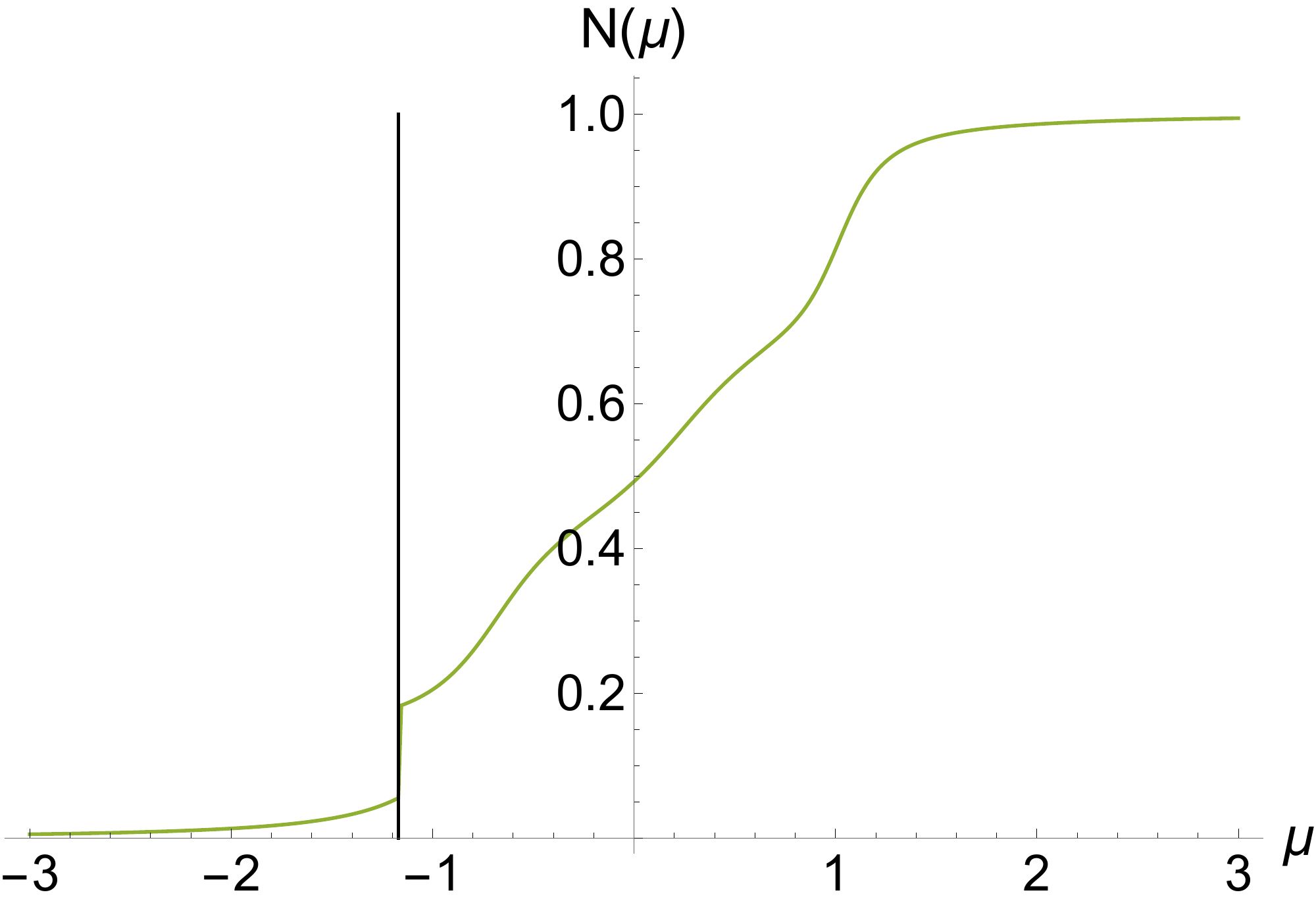} }
\end{minipage}
    \caption{The expectation value of the number operator as a function of $\mu$ with $J = 1, \Delta = \frac{1}{2}$, $a=1, b=2$.  We have fixed $\phi_1 = -\frac{\pi}{4}$.  Whenever the energy gap between the first and excited state closes we observe a discontinuity develop in the expectation value.  The values of $\mu$ where this happens are pointed out by a vertical black line which lines up exactly with the discontinuity.}
    \label{fig:ExpectationValueEx2}
\end{figure}

 In this appendix we would like to present some preliminary results related to the twisted Kitaev chain.  The Hamiltonian and site-dependent chemical potential of this model are given already in (\ref{TwistedKitaevH}). 
In order to compute the spread complexity for this model we first need to compute the Krylov basis.  For general parameters this is a non-trivial task. The boundary terms present in (\ref{TwistedKitaevH}) break translational invariance so that a Fourier transform no longer yields a tensor product structure.  As a result of this the eigenstates for the twisted Kitaev chain can no longer be expressed as $SU(2)$ coherent states.  Obtaining the Krylov basis is thus a more formidable task which we will circumvent by making a suitable approximation. \\ \\
The approximation we apply is based on the following considerations.   
In Appendix \ref{su2Appendix} we showed that the $SU(2)$  K-complexity is, up to a constant, the expectation value of $J_0$ with respect to an $SU(2)$ coherent state.  When performing the inverse Fourier transform on (\ref{su2Generators}) we thus find that the $SU(2)$ K-complexity is the expectation value of the number operator with respect to this coherent state.  Physically the number operator expectation value makes sense as a good measure of complexity since we need to add fermions to the vacuum in order to obtain the target states and adding more fermions is proportionally more costly in terms of circuit operations.  \\ \\
In light of this result we calculate the expectation value of the number operator as an estimate of the spread complexity.  We stress that this is not the spread complexity for the twisted boundary condition, since we have not computed the Krylov basis corresponding to the twisted Kitaev chain Hamiltonian.  We expect, however, that the number operator expectation value will capture similar features.  Explicitly, we compute the ground state of the twisted Kitaev chain and then compute the expectation value
\begin{equation}
N = \langle \psi_{gs}| \sum_{i=1}^L c_i^\dag c_i |\psi_{gs}\rangle    \label{NExpectationV}
\end{equation}
as our rough estimate of complexity.  Since we are working at finite $L$ we have to, in addition to computing the expectation value, compute the spectrum numerically and determine whether there is gap closing. As a cross-check of the accuracy of our numerical results we have compared with the analytical results of \cite{Kawabata:2017zsb} in the cases where exact results for finite $L$ are known. \\ \\
There are many parameters to adjust in the twisted Kitaev chain.  As such, our numerical studies probably do not cover all interesting features.  We do, however, report the following observation relevant to topological phase transitions:  Over a significant sample size of cases we have found that, typically, a closing energy gap between the ground state and first excited state (or equivalently, a degenerate ground state) can be identified by the expectation value (\ref{NExpectationV}). Specifically, these may be identified as discontinuities in the expectation value in the presence of the twisted boundary.  \\ \\
As an instructive example, consider the case $J = \Delta=1$ and $a=b=1$.  For these parameters and even $L$ we find, numerically, a simple relation for the existence of gap closing namely $|\phi_1| < |\phi_2|$ in the interval $\phi_i \in [-\pi, \pi)$.  By this we mean that there exist values for the chemical potential in the interval $|\mu|<1$ for which the ground state becomes degenerate.  Furthermore, for even $L$ we always observe an even number of points for which this occurs, see Fig. (\ref{fig:EnergyGap}) for $L=8$ as a representative example of this. \\ \\
\begin{figure}
\begin{minipage}{0.32 \textwidth}
    \centering
    \subfloat[$L=8, \phi_2 = \frac{\pi}{8}$]{\includegraphics[width=0.9\textwidth]{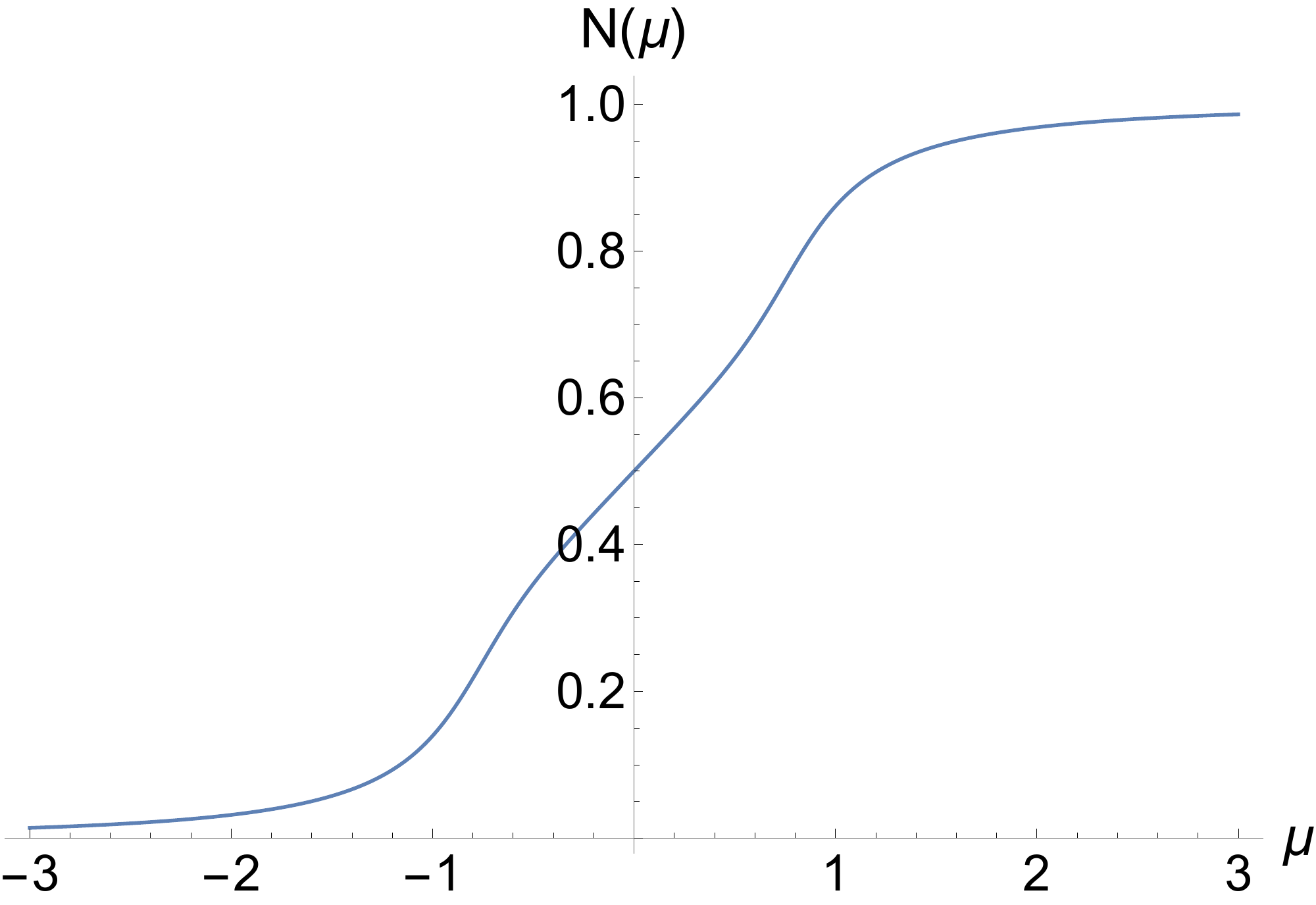} }
    \vfill
     \subfloat[$L= 9, \phi_2 = \frac{\pi}{8}$]{\includegraphics[width=0.9\textwidth]{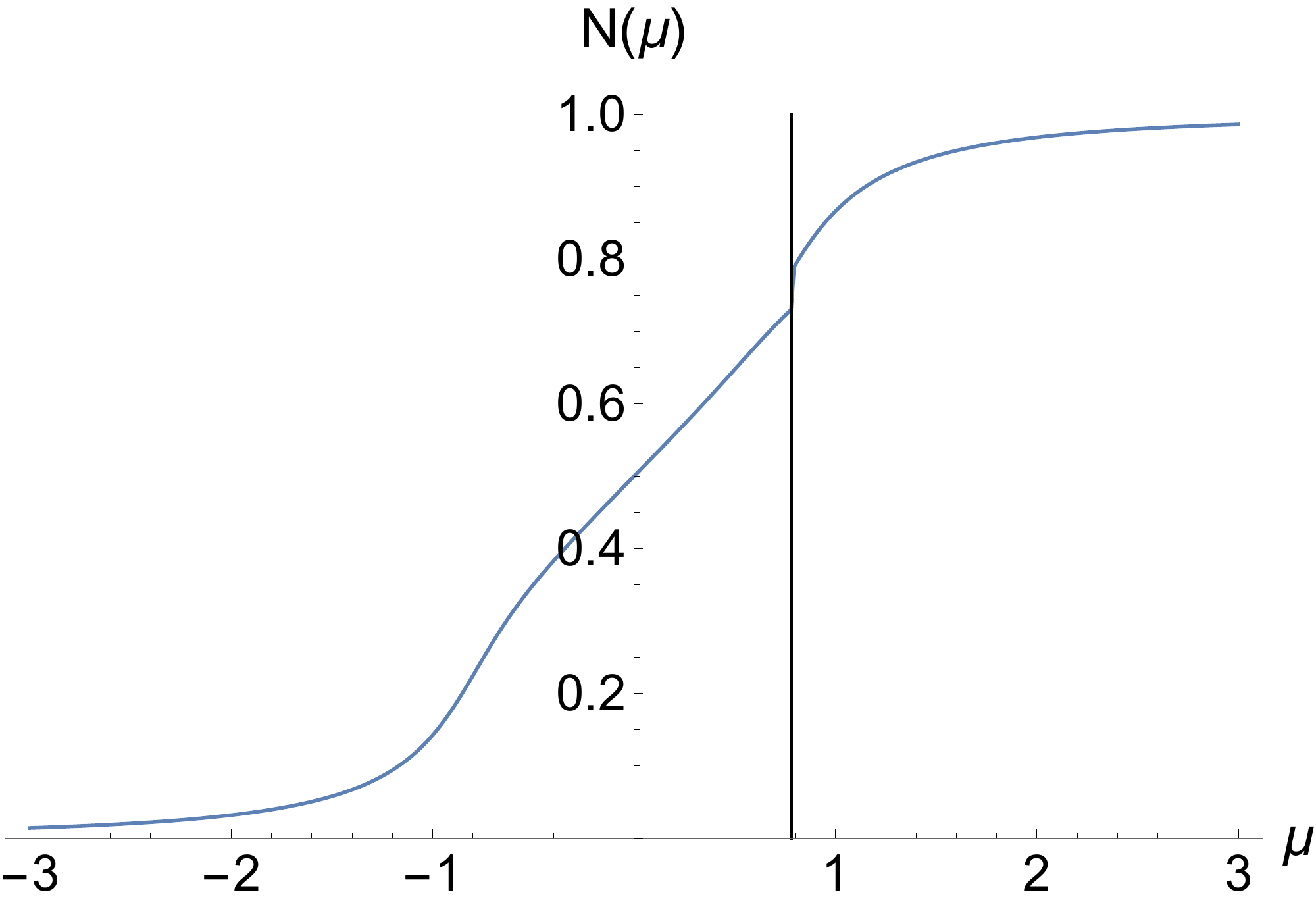} }
\end{minipage}
\begin{minipage}{0.32 \textwidth}
    \centering
    \subfloat[$L= 8, \phi_2 = \frac{\pi}{2}$]{\includegraphics[width=0.9\textwidth]{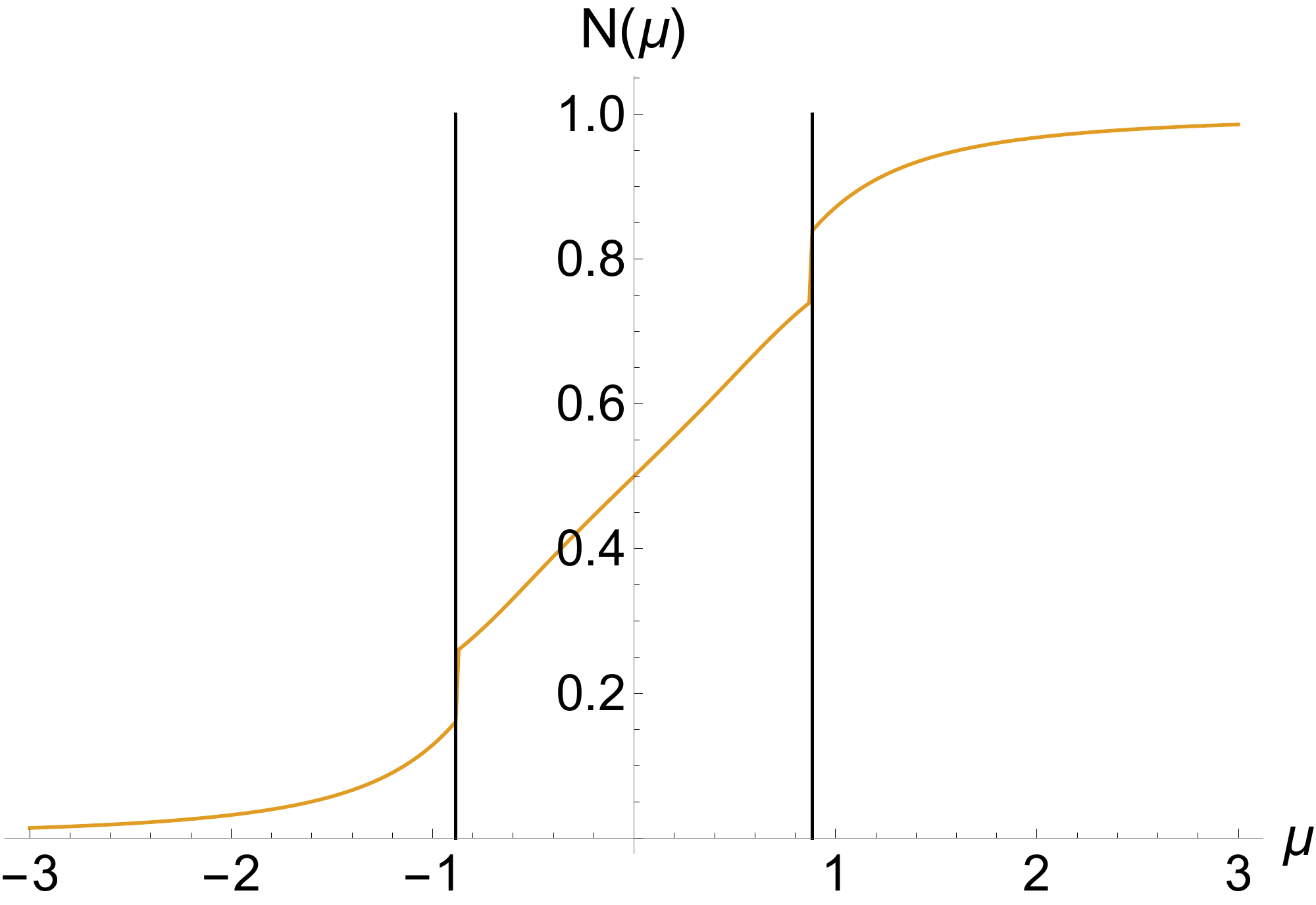} }
    \vfill
     \subfloat[$L= 9, \phi_2 = \frac{\pi}{2}$]{\includegraphics[width=0.9\textwidth]{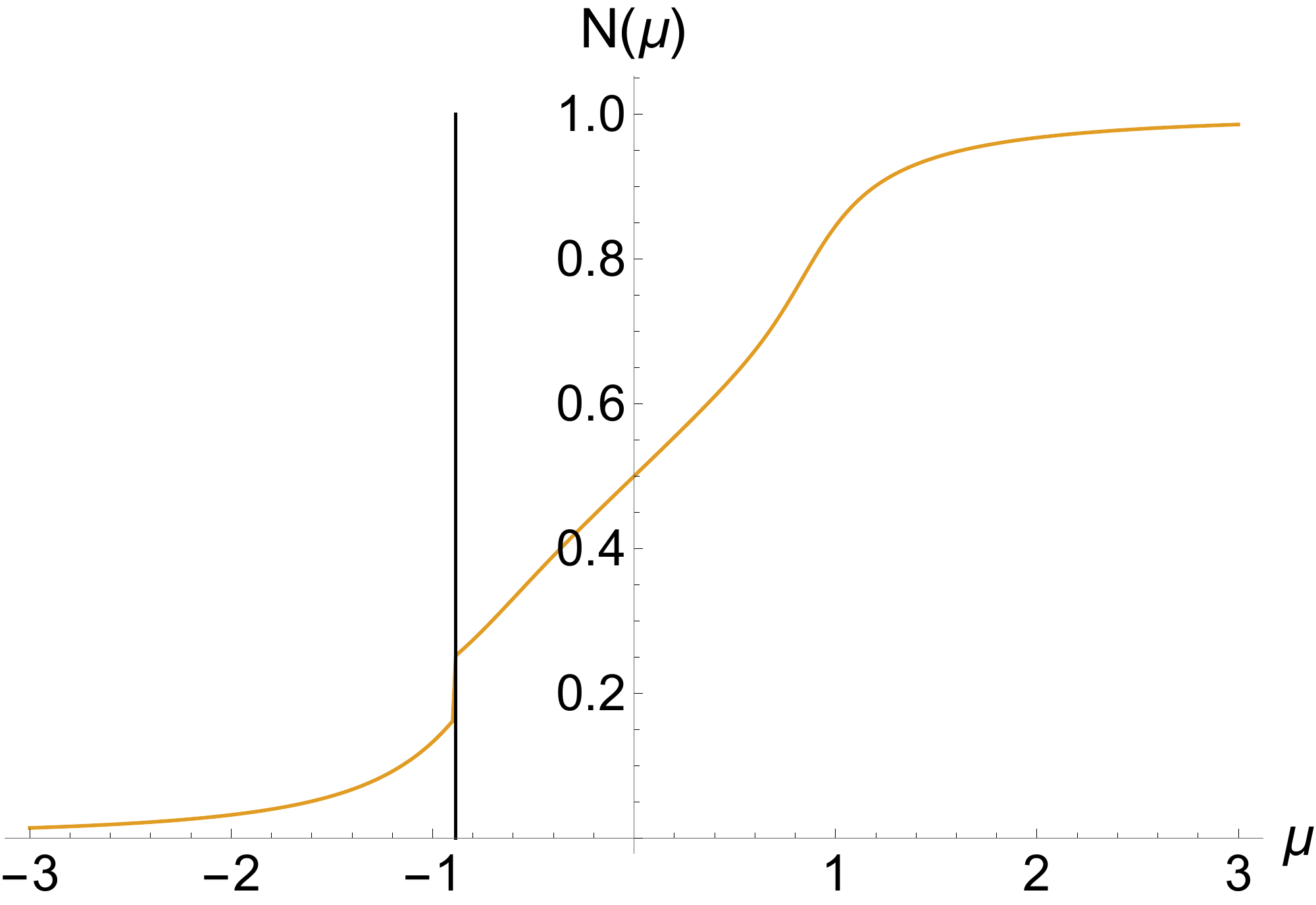} }
\end{minipage}
\begin{minipage}{0.32 \textwidth}
    \centering
    \subfloat[$L=8, \phi_2 = \frac{7\pi}{8}$]{\includegraphics[width=0.9\textwidth]{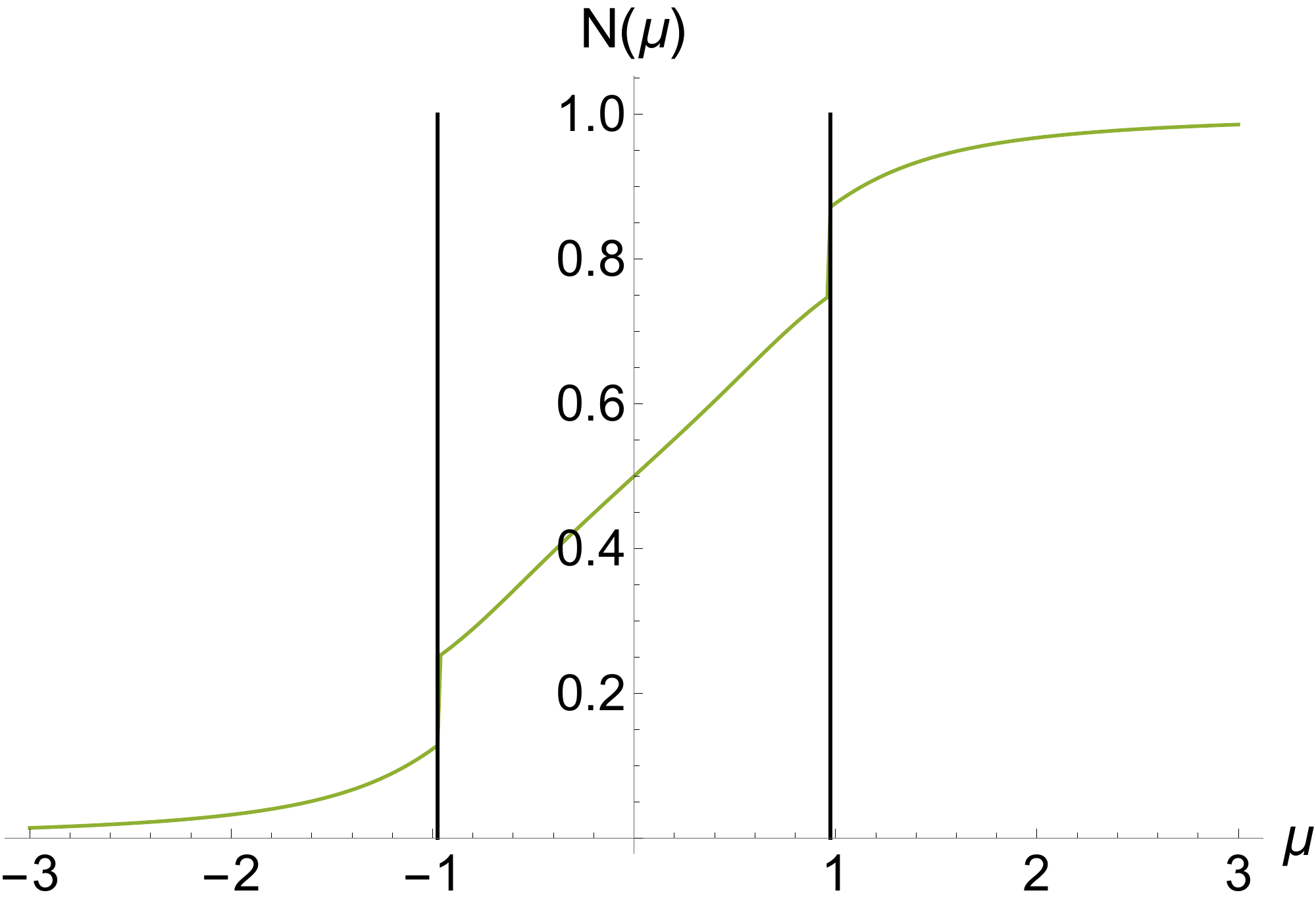} }
    \vfill
     \subfloat[$L= 9, \phi_2 = \frac{7\pi}{8}$]{\includegraphics[width=0.9\textwidth]{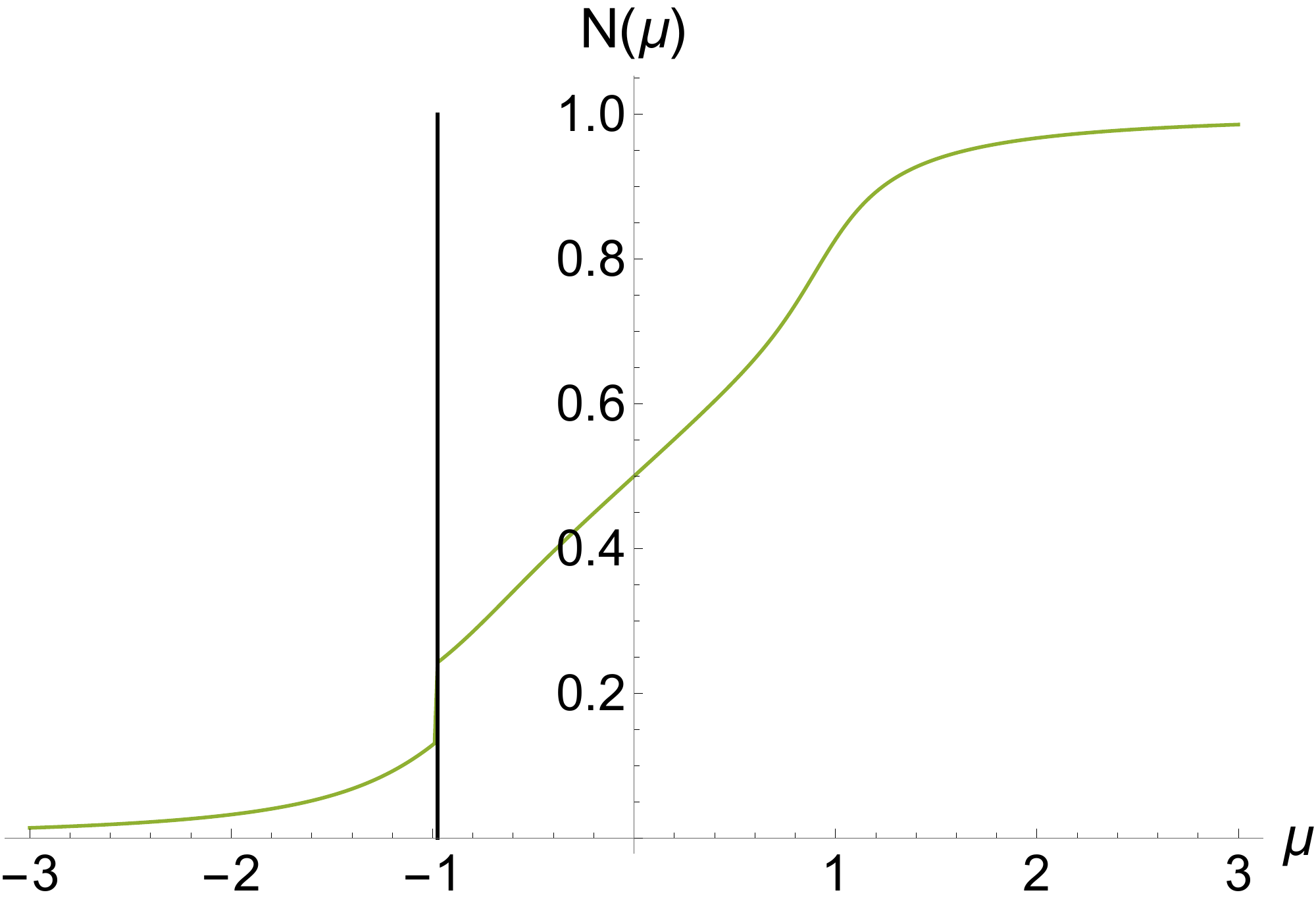} }
\end{minipage}
    \caption{The expectation value of the number operator as a function of $\mu$ with $\Delta = J = 1$, $a=b=1$.  We have fixed $\phi_1 = -\frac{\pi}{4}$.  Whenever the energy gap between the first and excited state closes we observe a discontinuity in the expectation value.  The values of $\mu$ where this happens are pointed out by a vertical black line which lines up exactly with the discontinuity.}
    \label{fig:ExpectationValue}
\end{figure}
For odd $L$ we usually find an odd number of values of $\mu$ where this occurs and it can also occur for some values where $|\phi_2| < |\phi_1| $. 
For both even and odd $L$ cases, see Fig. (\ref{fig:ExpectationValue}), we observe the advertised discontinuity whenever the ground state becomes degenerate. The larger the difference $|\phi_2| - |\phi_1|$ the bigger the discontinuity. \\ \\
As a second example consider the case $J=1, \Delta = \frac{1}{2}$ and $a=1, b=2$.  We highlight this case for a two reasons.  Firstly, in the case of even $L$ we no longer find the simple relation $|\phi_1| < |\phi_2|$ to predict the existence of gap closing at some value of $\mu$.  Secondly, we see the appearance of a third location where this occurs in Fig. (\ref{fig:ExpectationValueEx2}) (b).  In general we find an even (odd) number of values of $\mu$ for which gap closing can occur for even (odd) $L$.  Both these observation resonate with the analytical results of \cite{Kawabata:2017zsb} where an even (odd) number of zero modes can be found for a chain with an even (odd) number of sites and, in general, the equations determining the existence of zero modes are involved for finite $L$. \\ \\
The intuition for the observed discontinuity in (\ref{NExpectationV}) is rather straightforward.  As we are changing the value of $\mu$ continuously the spectrum and eigenstates change continuously.  Focusing on one of these states, the expectation value of the number operator w.r.t this state will also change continuously.  However, when we encounter a value of $\mu$ where the ground state energy gap closes, there is a discontinuous change in the ground state wave function since the ground state and first excited state wave functions are exchanged in the spectrum.  This accounts for the discontinuities we observe in the number operator expectation value. 
\begin{figure}[!h]
\begin{subfigure}{.49\textwidth}
	\centering
		\includegraphics[width=0.90\textwidth]{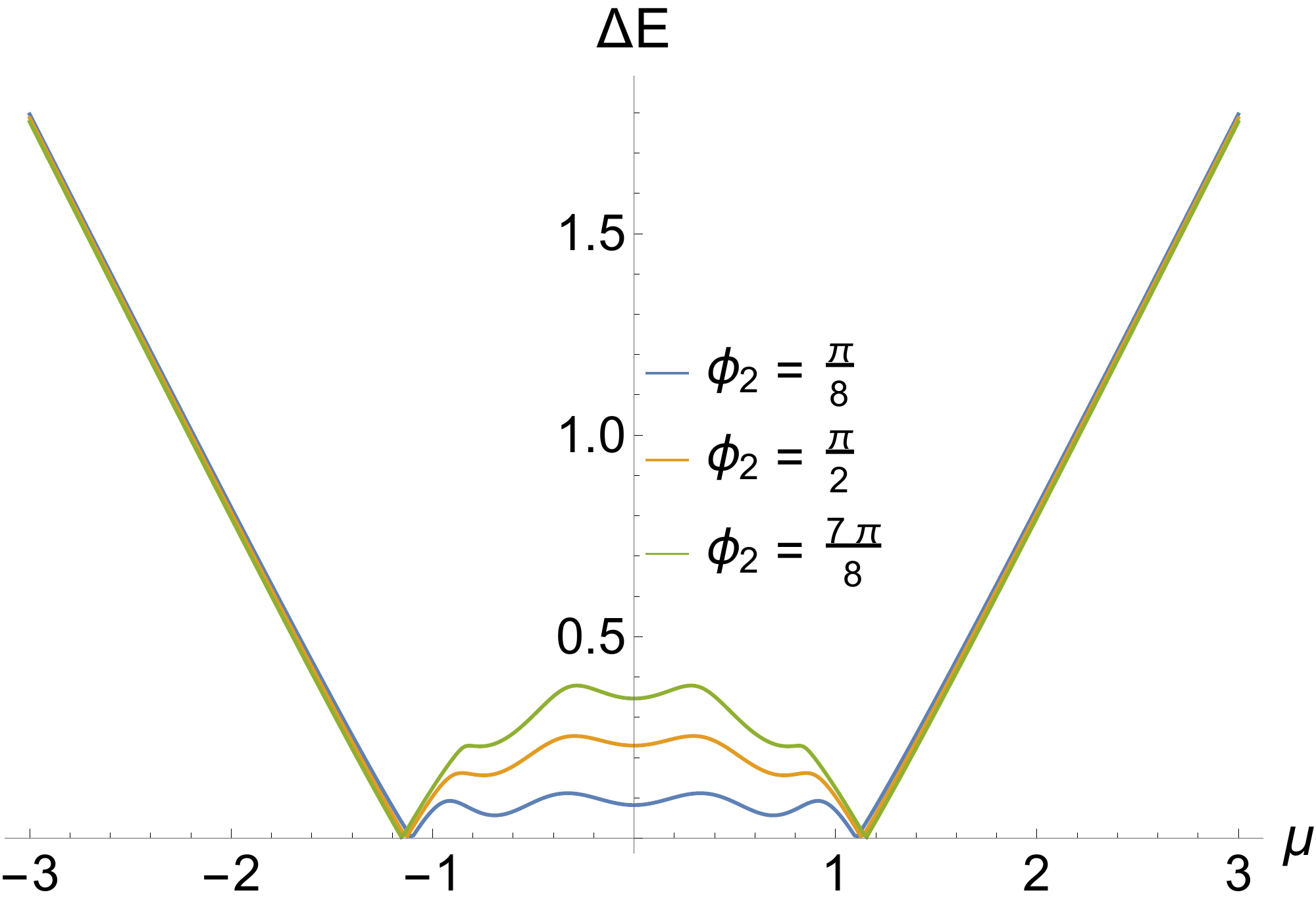}
	\caption{ }
	\label{fig:spreadComplexityTwist2}
	\end{subfigure}
	\begin{subfigure}{.49\textwidth}
		\centering
		\includegraphics[width=0.90\textwidth]{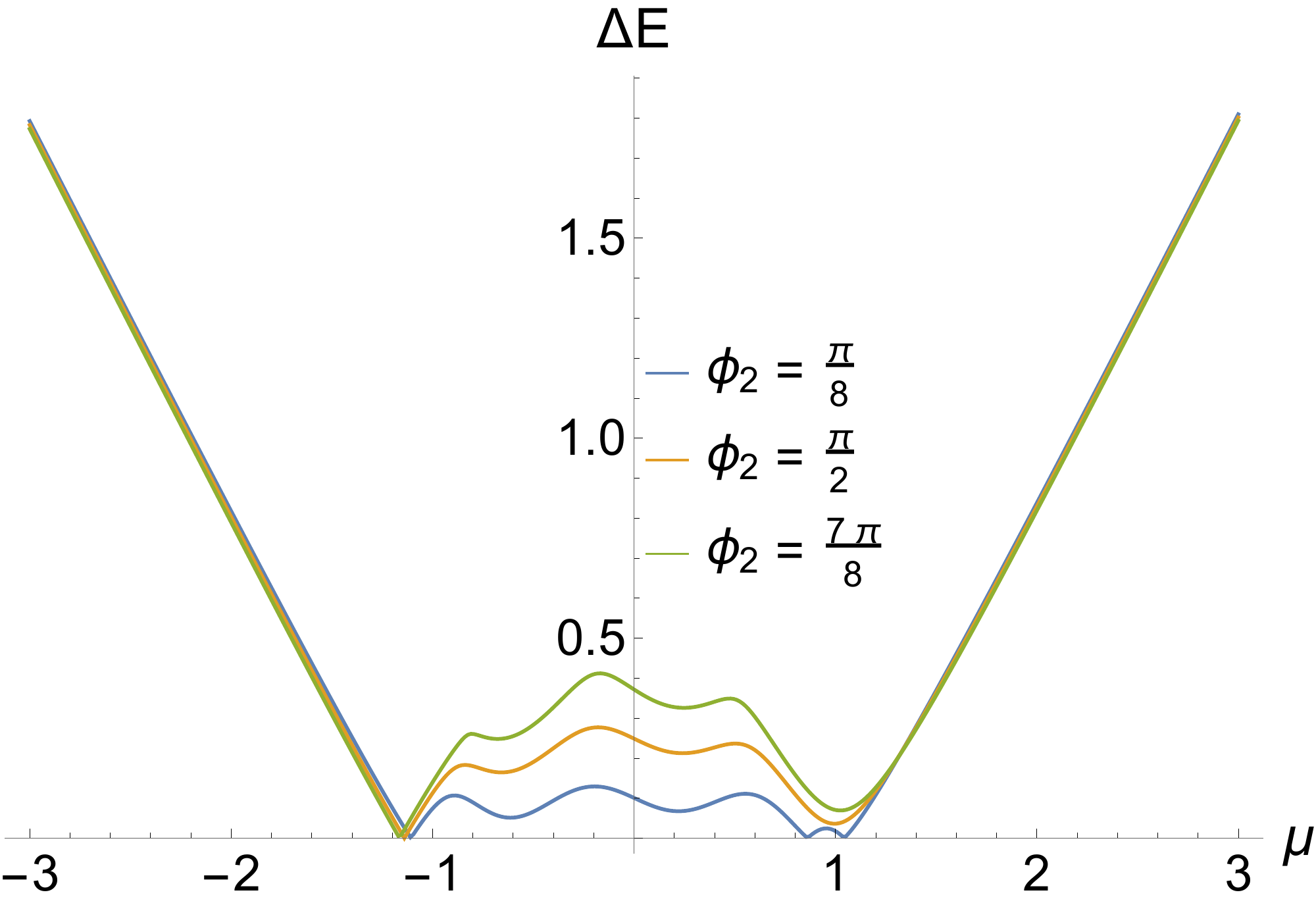}
	\caption{ }
	\label{fig:EnergyGapEx2}
	\end{subfigure}
	\caption{ The energy gap between the ground state and first excited as a function of $\mu$ with $J=1, \Delta = \frac{1}{2}$, $a=1, b=2$.  We have fixed $\phi_1 = -\frac{ \pi}{4}$.   a) The chain length is $L=8$.  For these parameters we note a pair of values for $\mu$ exhibit a closing energy gap provided $|\phi_2| > |\phi_1|$    b) The chain length is $L=7$. In this case we observe only a single value for $\mu$ for which gap closing occurs for $|\phi_2| > |\phi_1|$.  We find three values for which this occurs for $|\phi_2| < |\phi_1|$  } 
\end{figure}

\noindent
To summarize; in the context of the twisted Kitaev chain the sensitivity of spread complexity to the topological phase transition is related to the sensitivity of the number operator expectation value to the ground state degeneracy.  A degenerate ground state signals a discontinuous change in the expectation value, albeit at finite $L$, due to a discontinuous change in the ground state wavefunction. The spread complexity, in the large $L$ limit, is precisely the expectation value of the number operator with respect to some eigenstate of the Hamiltonian.  At finite $L$ this connection is not exact, since the Krylov basis should be updated in the presence of the twisted boundaries. Nevertheless, we expect this connection to be a good approximation to the true spread complexity. \\ \\

\end{document}